\documentclass[pdflatex,iicol,sn-mathphys-num]{sn-jnl}
\usepackage{graphicx}
\usepackage{amsmath}
\usepackage{amssymb}
\usepackage{bm}
\usepackage{array}
\usepackage{booktabs}
\usepackage[table]{xcolor}
\usepackage[english]{babel}

\newcommand{\topcaption}{\caption}

\newcommand{\nPapersAtlas}{28}
\newcommand{\nPlotsAtlas}{168}

\newcommand{\nPapersCms}{27}
\newcommand{\nPlotsCms}{205}

\newcommand{\nPapers}{55}

\newcommand{\resRawOneAtlas}{81.8 \pm 5.9}
\newcommand{\devRawOneAtlas}{2.3}
\newcommand{\resRawTwoAtlas}{99.5 \pm 0.3}

\newcommand{\resNormOneAtlas}{73.3 \pm 7.6}
\newcommand{\devNormOneAtlas}{0.7}
\newcommand{\resNormTwoAtlas}{99.2 \pm 0.4}

\newcommand{\resRawOneCms}{69.5 \pm 3.2}

\newcommand{\resRawTwoCms}{95.8 \pm 1.2}

\newcommand{\resNormOneCms}{66.1 \pm 4.4}

\newcommand{\resNormTwoCms}{95.3 \pm 1.9}

\newcommand{\tsRawOneAtlasEarly}{87.0 \pm 6.1}
\newcommand{\tsRawOneAtlasLate}{75.9 \pm 8.6}

\newcommand{\tsDiffMaxAtlas}{1.8}
\newcommand{\tsRawOneCmsEarly}{72.7 \pm 3.8}
\newcommand{\tsRawOneCmsLate}{65.3 \pm 5.3}

\newcommand{\tsDiffMinCms}{1.1}
\newcommand{\tsDiffMaxCms}{2.1}
\newcommand{\methMaxSig}{2.1}
\newcommand{\methMaxWhat}{CMS raw $1\sigma$}
\newcommand{\methMaxDiff}{+3.8 \pm 1.8}
\newcommand{\looUpRawOneAtlas}{3.7}
\newcommand{\looUpPubRawOneAtlas}{arXiv:2305.15962}
\newcommand{\looDownRawOneAtlas}{4.1}
\newcommand{\looDownPubRawOneAtlas}{arXiv:2201.07045}
\newcommand{\looUpNormOneAtlas}{7.6}
\newcommand{\looUpPubNormOneAtlas}{arXiv:2305.15962}

\newcommand{\looUpRawOneCms}{1.4}

\newcommand{\looDownRawOneCms}{1.2}

\newcommand{\looUpNormOneCms}{1.8}

\newcommand{\looDownNormOneCms}{2.7}

\newcommand{\pixComparedCol}{19128}

\newcommand{\pixEdgeCol}{4292}

\newcommand{\pixComparedPoly}{40765}

\newcommand{\pixEdgePoly}{8648}

\newcommand{\clusterPAtlas}{= 0.0016}
\newcommand{\clusterPerm}{20000}
\newcommand{\uncBinomAAtlas}{3.9}
\newcommand{\uncPlotsAAtlas}{4.1}
\newcommand{\uncBinomAtlas}{2.9}
\newcommand{\uncGroupsAtlas}{3.9}
\newcommand{\uncPubsAtlas}{6.3}
\newcommand{\clusterGroupsAtlas}{64}
\newcommand{\clusterPubsAtlas}{28}
\newcommand{\clusterPCms}{= 0.000050}
\newcommand{\uncBinomACms}{3.4}
\newcommand{\uncPlotsACms}{3.4}
\newcommand{\uncBinomCms}{2.8}
\newcommand{\uncGroupsCms}{3.1}
\newcommand{\uncPubsCms}{4.4}
\newcommand{\clusterGroupsCms}{104}
\newcommand{\clusterPubsCms}{27}
\newcommand{\citePAllAtlas}{= 0.33}
\newcommand{\citePairsAtlas}{33}
\newcommand{\citeLinkedAtlas}{19}
\newcommand{\citePSepAtlas}{= 0.18}
\newcommand{\citePAllCms}{= 0.69}
\newcommand{\citePairsCms}{18}
\newcommand{\citeLinkedCms}{18}
\newcommand{\citePSepCms}{= 0.83}

\begin{document}

\title[Updated meta-analysis of LHC exotic physics results]{An Updated Meta-Analysis of LHC Exotic Physics Results}

\author*[1]{\fnm{Ece} \sur{\c{S}ahin}}\email{ece.sahin@std.bogazici.edu.tr}
\author[2,3,4]{\fnm{Veysi Erkcan} \sur{\"Ozcan}}

\affil*[1]{\orgdiv{Department of Physics}, \orgname{Bo\u{g}azi\c{c}i University}, \orgaddress{\city{Istanbul}, \country{T\"urkiye}}}
\affil[2]{\orgdiv{Department of Basic Sciences}, \orgname{\.{I}stinye University}, \orgaddress{\city{Istanbul}, \country{T\"urkiye}}}
\affil[3]{\orgdiv{Feza G\"ursey Center for Physics \& Mathematics}, \orgname{Bo\u{g}azi\c{c}i University}, \orgaddress{\city{Istanbul}, \country{T\"urkiye}}}
\affil[4]{\orgdiv{Center for Targeted Therapy Technologies}, \orgname{Bo\u{g}azi\c{c}i University}, \orgaddress{\city{Istanbul}, \country{T\"urkiye}}}

\abstract{The replication crisis has raised concern across many fields of science about whether published results are as reliable as their stated uncertainties imply. Searches for new physics at the Large Hadron Collider offer a direct test of this for high-energy physics: those that find no signal report exclusion limits together with the limits expected in the absence of a signal and their $1\sigma$ and $2\sigma$ bands, which readers take to contain the observed limits with frequencies of 68.3\% and 95.4\%. Following a preliminary study in 2019, we test this reading for the ATLAS and CMS searches published between 2020 and 2025, 55 publications with nearly 400 exclusion plots, and reanalyze the publications of the earlier study. The limits are read directly from the vector graphics of the figures, and the statistical uncertainties are estimated with a bootstrap over publications. Averaging over the plots with equal weights, the CMS results agree with that reading, with $66.1 \pm 4.4\%$ and $95.3 \pm 1.9\%$ of the observed points within the $1\sigma$ and $2\sigma$ bands. For ATLAS, $73.3 \pm 7.6\%$ of the points lie within the $1\sigma$ band, compatible with it, while $99.2 \pm 0.4\%$ lie within the $2\sigma$ band, more often than it implies. Pseudo-experiments show that this can arise from conservative uncertainties, from the few events of many searches, or from expected limits computed after a fit to the data, the default of common statistical tools, which the plots do not distinguish. We recommend that publications state how their expected limits are computed.}

\keywords{LHC, ATLAS, CMS, meta-analysis}

\maketitle

\section{Introduction}
The replication crisis refers to the growing number of published studies whose results fail to be reproduced when independent researchers attempt to replicate them~\cite{ioannidis2005,baker2016}. The question we address is related but narrower: not whether results can be reproduced, but whether the uncertainty bands that accompany published exclusion limits are calibrated, that is, whether they contain the observed limits as often as their nominal probabilities say. Experimental high energy physics is generally considered to be less prone to this issue due to its strong culture of blind analysis and open data, as well as the use of large datasets, which reduce the probability of potential biases. Despite these advantages, it remains important to regularly verify whether any unexpected biases, however small, get introduced as the statistical tools get updated, new samples are collected, systematic uncertainties change as the detectors age.

The Large Hadron Collider experiments have published hundreds of articles on the searches for new physics beyond the Standard Model (BSM). While no positive discovery results have been reported yet, these studies provide constraints on the masses of hypothetical particles, their production cross-sections, and various other parameters such as the coupling constants. These constraints are often summarized as exclusion limits, the extraction of which requires substantial statistical modeling and a detailed understanding of the behavior of the detectors. Accordingly, LHC collaborations organize dedicated subgroups to establish common policies on which statistical analysis libraries to use and how to treat systematic uncertainties across different sub-detectors. In this context, examining the consistency and reliability of the exclusion analyses represents a straightforward means for identifying possible indications of a replication problem.

In a previous study published in 2019, we performed an analysis focusing on possible biases in the reported uncertainty ranges in papers from ATLAS and CMS experiments~\cite{aciksoz2019}. We calculated the fraction of data points falling inside the $1\sigma$ and $2\sigma$ bands and compared those fractions to the statistical coverage expected under unbiased uncertainty estimates. In this preceding study, the $2\sigma$ results matched the expected 95\% coverage for both experiments. The CMS $1\sigma$ result, however, was higher than the expected 68\%, while ATLAS agreed.

At the time of the 2019 study, the numerical results behind only a few of these plots were available in HEPData. We therefore read the plots themselves: the figures of the publications are vector graphics, which we convert to the SVG format, so that the observed limit and the uncertainty bands can be extracted from their geometry exactly, without digitizing an image. Although the number of HEPData entries has grown since, its coverage remains incomplete (Section~\ref{sec:hepdata}), and the present study follows the same approach.

The present study applies this method to the ATLAS and CMS publications of 2020--2025 and revisits the publications of the 2019 study with the updated code. The method has been extended to plot types that previously had to be excluded, such as limit curves that loop back onto themselves and binned plots, and the statistical uncertainties of the results are now estimated with a bootstrap over publications.

As before, the aim is to test whether the reported uncertainty bands have their nominal coverage, and thereby to look for indications of systematic biases in how exclusion limits are derived.

The structure of this manuscript is as follows. Section~\ref{sec:data} describes the dataset. Section~\ref{sec:method} outlines the methodology. Section~\ref{sec:results} presents the results, followed by consistency checks in Section~\ref{sec:validation}. Finally, Section~\ref{sec:conclusions} concludes the paper.

\section{Data}\label{sec:data}
We utilize exclusion plots, with an example shown in Fig.~\ref{sceleton}, taken from the public results pages of the corresponding experiments. These plots visualize theoretical predictions that are disfavored by the experimental data. More concretely, they denote the expected and observed upper limits on the signal production cross section as a function of an experimentally reconstructed variable, such as the mass of a hypothetical particle. 

The observed limit is commonly plotted with a black solid curve. Regions above the observed curve at a specified confidence level are considered to be inconsistent with the experimental findings and excluded from the allowed parameter space. Additionally, one also observes a dashed black line, indicating the expected limit: the median limit expected in the absence of a signal. It is obtained from pseudo-experiments or, more often, from a representative (Asimov) data set with asymptotic formulae, and either a priori, from the nominal background expectation, or a posteriori, from a background-only fit to the observed data.

$1\sigma$ and $2\sigma$ bands, which are generally plotted as green and yellow regions respectively, indicate the uncertainty on the exclusion limit. Readers, including those who reinterpret the results, take the bands as the intervals that contain the observed limit 68.3\% and 95.4\% of the time if no signal is present. That is what they are if the expected limits are computed a priori, from the nominal background expectation, and the asymptotic formulae apply. Other conventions and approximations change what the bands contain (Section~\ref{sec:interpretation}).

The observed curves are not smooth: the limits are computed at the parameter values for which signal templates exist, the benchmark points, and the curves are drawn as straight lines between them. This discreteness is usually distinguishable in the plots, as the benchmark points appear as kinks in the curves, often in the boundaries of the uncertainty bands as well. Our analysis therefore reads the observed limit only at the vertices of its curve and discards the segments drawn between them; the column scan, our main method, in addition accepts only those vertices at which the band boundaries have vertices too (Section~\ref{sec:method}).

All exclusion plots processed in this study are collected from a total of \nPapers{} papers, \nPapersAtlas{} from ATLAS and \nPapersCms{} from CMS\footnote {The reasons for excluding certain plots are explained in Section~\ref{sec:excluded}. In some cases, the exclusion of all relevant plots from a given paper resulted in that paper being omitted entirely; such papers are therefore neither counted nor listed in Table~\ref{tab:papers}.}, the arXiv identifiers of which are listed in Table~\ref{tab:papers}. Following our previous study, we include only papers published within the last five years, selected by predetermined topics of exotic physics\footnote{For CMS, papers under the exotic physics theme are further classified into excited fermions, leptoquarks, heavy additional fermions, compositeness, extra dimensions, and contact interactions.\\For ATLAS, BSM search papers are restricted to vector-like quarks, leptoquarks, heavy neutrinos, compositeness, and contact interactions.}.

\begin{table*}[tb]
\centering
\topcaption{arXiv identifiers of the papers from which the exclusion plots are obtained.}\label{tab:papers}
% generated by analysis/papertables.py from base git 2c7a2e8, 2026-11-03; do not edit
\begin{tabular}{ccc@{\hspace{2.5em}}c@{\hspace{2.5em}}ccc}
\toprule
\multicolumn{3}{c}{ATLAS} & & \multicolumn{3}{c}{CMS} \\
\midrule
2503.22581 & 2305.15962 & 2202.02039 &  & 2506.13565 & 2308.06143 & 2202.06075 \\
2503.19836 & 2305.14931 & 2201.07045 &  & 2503.20023 & 2307.08708 & 2201.05578 \\
2409.20273 & 2305.03401 & 2105.13847 &  & 2410.21137 & 2307.06959 & 2112.03949 \\
2405.19862 & 2304.09553 & 2101.12527 &  & 2407.10717 & 2305.07998 & 2107.13021 \\
2403.15016 & 2303.09444 & 2010.02098 &  & 2405.09320 & 2212.12604 & 2012.04178 \\
2403.06742 & 2303.01294 & 2008.07949 &  & 2402.18658 & 2210.03082 & 2008.04735 \\
2402.16561 & 2212.05263 & 2006.12946 &  & 2402.08668 & 2206.08956 & 2001.04521 \\
2401.17165 & 2210.15413 & 2006.05872 &  & 2312.07484 & 2205.06709 &  \\
2401.11928 & 2210.04517 &  &  & 2310.14023 & 2205.01835 &  \\
2307.07584 & 2204.11988 &  &  & 2308.07826 & 2202.08676 &  \\
\bottomrule
\end{tabular}

\end{table*}

\subsection{Dataset Scope and Selection Criteria}\label{sec:scope}
While aiming to include all exclusion plots from the listed papers, we encountered several problematic cases. Some of these could be addressed using newly developed algorithms, whereas others had to be excluded due to structural limitations. In addition, certain types of plots could not be analyzed in the previous study because of their non-standard structures. In the present analysis, dedicated algorithms have been developed to handle these cases, thereby extending the scope of the dataset. This subsection discusses the resulting limitations and extensions of the dataset. 

\subsubsection{Plots Excluded from the Analysis}\label{sec:excluded}
Our analysis method is designed to process exclusion plots under a set of well-defined structural and statistical requirements. Therefore, several classes of plots that do not satisfy one or more of these requirements had to be excluded from the analysis. 

\begin{figure*}[tb]
\centering
\includegraphics[height=5.6cm]{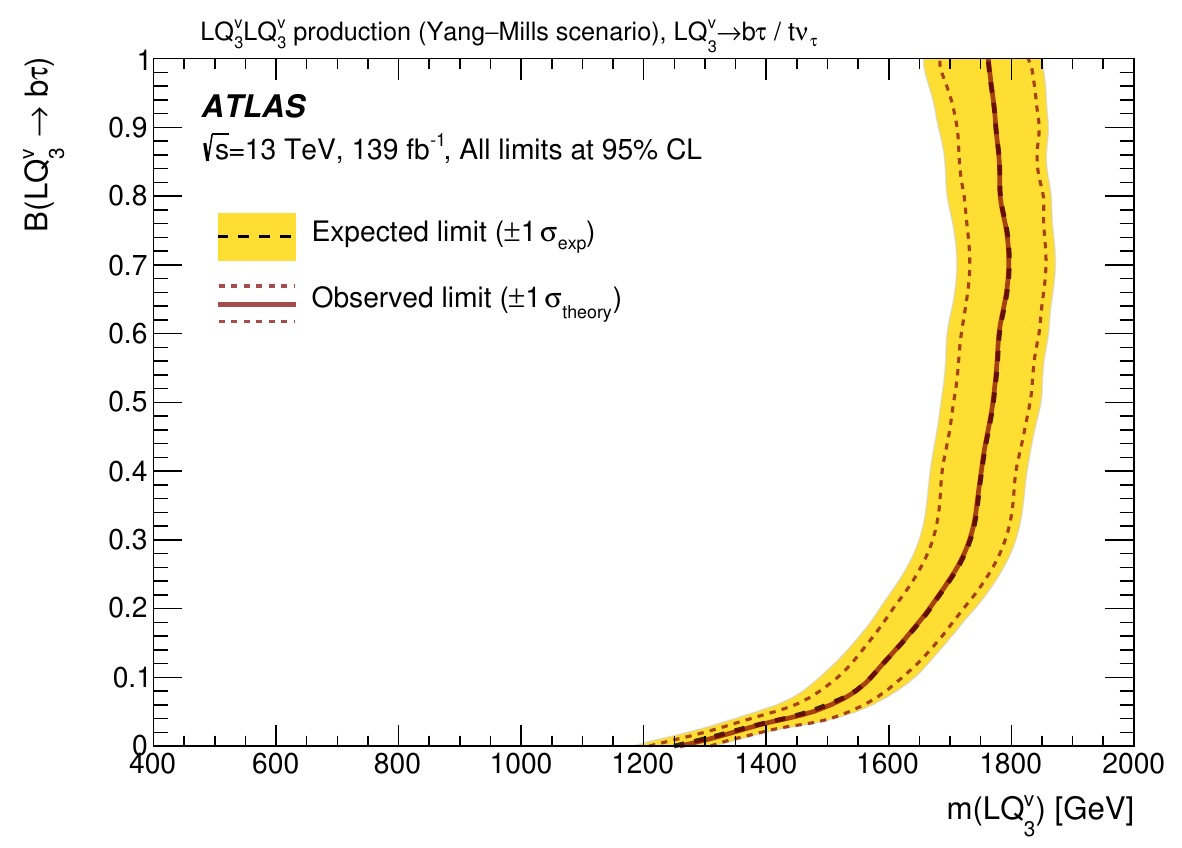}\hfill
\includegraphics[height=5.6cm]{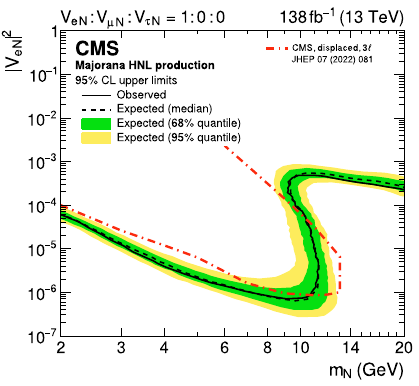}
\caption{Two kinds of excluded plots: (a) a plot from arXiv:2108.07665~\cite{atlas_2108_07665} with only a $1\sigma$ band; (b) a ``snake'' plot from arXiv:2312.07484~\cite{cms_2312_07484} (Fig.~6a), in which the limit bends back on itself, so that a vertical line meets the observed curve and the bands more than once.}
\label{fig:onesigma}
\end{figure*}

Some exclusion plots are set aside when they are collected, each with its reason recorded alongside the data. In the present collection, the following plots are excluded: (i)~38 plots that show only a single uncertainty band (e.g., Fig.~\ref{fig:onesigma}a); (ii)~9 plots embedded in the PDF files as raster images, which contain no vector paths to analyze; (iii)~7 figures that duplicate another figure of the same paper; (iv)~11 so-called snake plots, whose limit curves bend back on themselves in an S shape (Fig.~\ref{fig:onesigma}b), so that the observed curve and the bands cannot be compared column by column; (v)~11 plots that show no exclusion limit at all (distributions of $p$-values and mass spectra); and (vi)~two plots that show several overlapping limit curves at once. These exclusions depend on how a plot is drawn, not on where its observed limit lies relative to its bands. We also do not use the 26 plots of the ATLAS summary paper arXiv:2403.09292: it re-presents results of other publications in our collection, so that its plots are not independent of theirs, and the paper does not appear in Table~\ref{tab:papers}.

Other plots are excluded automatically during the analysis, or, where the analysis finds no usable point in any plot of a paper, together with that paper. A plot is not used if none of its observed points can be matched to the band boundaries (Fig.~\ref{nogoodpts}, Section~\ref{sec:method}). Figures that reappear identically elsewhere, for instance as auxiliary figures, are recognized by a geometric fingerprint and counted once (Section~\ref{sec:averaging}).

Finally, plots are excluded if there are indicators that the points they show are unlikely to have come directly from the analyses. Some plots draw both the observed curve and the bands with closely packed vertices, a few tenths of an SVG unit apart, like curves sampled at the resolution of a raster image. Such plots arise from contours of a fine grid, or from smoothed or interpolated curves. In such a case we cannot tell which of the drawn points are benchmark points and which are interpolation fillers. We therefore drop a plot if more than 95\% of the distinct $x$ positions of its accepted points have another one within $1/452$ of the width of its plotting frame (one SVG unit for a frame of standard size; see Section~\ref{sec:thresholds}). All 20 collected plots of ATLAS arXiv:2503.16213, whose observed limits are closed contours drawn in this way, are excluded by this criterion, so that this paper does not appear in Table~\ref{tab:papers}. Its HEPData record does not help either: it gives each curve, the observed and the expected limit and each band edge, as a separate contour of over a thousand points on its own fine grid of masses, not the limits at the analyzed mass values. Plots in which only the observed limit is drawn densely are kept: the fine mass scans of the diphoton resonance searches arXiv:1606.03833 and arXiv:1707.04147 of the 2019 collection have their bands drawn at regularly spaced mass values, and the points accepted there are well separated. The opposite case is ATLAS arXiv:2506.15515, whose band outlines have a vertex every 0.65~SVG units while its observed curve is drawn coarsely. No column then contains exactly two boundary points of each band, so no point can be used, and this paper does not appear in Table~\ref{tab:papers} either.

\subsubsection{Included Plots Exhibiting Irregular Features}\label{sec:irregular}
During the collection of all exclusion plots, we encountered a limited number of non-ideal cases. While these cases do not pose a challenge for the analysis algorithms, they are not optimal from the perspective of scientific presentation and standardization.

Uncertainty bands in exclusion plots are conventionally plotted in green and yellow. In the 2019 study, every ATLAS paper used one and the same pair of colors (pure green and pure yellow), whereas CMS used several shades. This is no longer the case. The ATLAS papers of the present collection use at least seven different pairs of band colors, nearly as many as the CMS papers (eight, including the scheme described next). Moreover, the only plots in our collection with the roles of the two colors inverted are from ATLAS (see below). CMS has further deviated from the conventional color scheme by using yellow and blue to represent the $1\sigma$ and $2\sigma$ bands, respectively (e.g., arXiv:2503.20023). Such plots are analyzed with the corresponding color assignment: when a plot contains no green band, its golden-yellow and sky-blue fills are taken as the $1\sigma$ and $2\sigma$ bands.

We do not, however, follow every deviation. In ATLAS arXiv:2507.03650 the roles of the two colors are inverted, a pale yellow band showing the $1\sigma$ and a green band the $2\sigma$ uncertainty, and both bands are drawn semi-transparently. Its plots are not used in this analysis, and the paper does not appear in Table~\ref{tab:papers}.

The observed limit, too, is not always drawn as a plain black line. Some plots draw it in red. Some mark the analyzed points with small dots, on the observed or on the expected limit, or add short ticks on the excluded side of the observed line (e.g., arXiv:2407.10717). And theory predictions often appear as colored lines with an uncertainty band of their own color. The rules by which the observed limit is identified in such cases are given in Section~\ref{sec:method}.

Another issue arises when the color scheme used in the legend does not match the colors used in the plot itself. In such cases, the stroke of the limit curves matches the legend, while the fill color differs from it. For instance, in arXiv:2503.16213, which is not part of the analysis, the $1\sigma$ band is filled with rgb(16\%, 99\%, 18\%), while its legend entry is rgb(0\%, 100\%, 0\%). This inconsistency complicates the interpretation of the uncertainty bands and represents a graphical issue that should not occur in standardized scientific visualizations.

\subsubsection{New Algorithms for Previously Unanalyzable Plot Types}\label{sec:newalgos}

\begin{figure*}[tb]
\centering
\includegraphics[height=5.6cm]{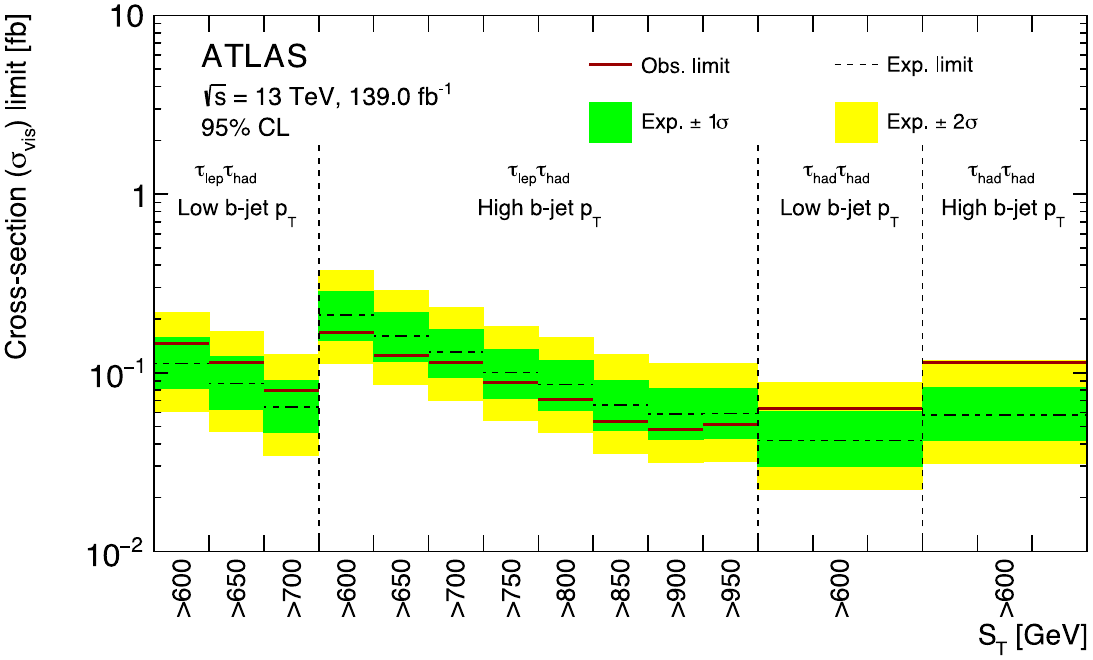}\hfill
\includegraphics[height=5.6cm]{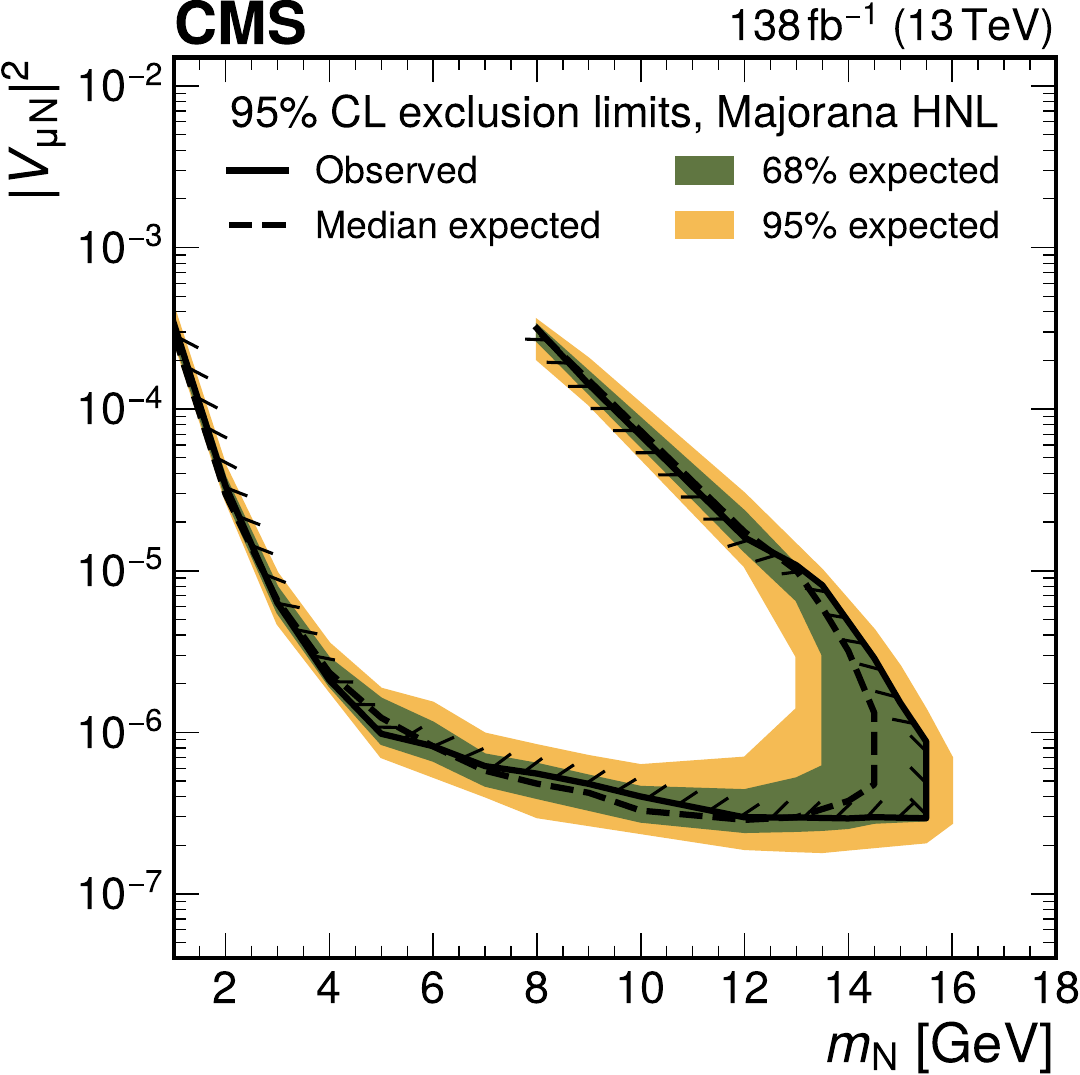}
\caption{Exclusion plots exhibiting non-regular structural characteristics, taken from arXiv:2305.15962~\cite{atlas_2305_15962} (Fig.~11) and arXiv:2407.10717~\cite{cms_2407_10717} (Fig.~13c).}
\label{unregularplots}
\end{figure*}

\begin{figure*}[tb]
    \centering
    \includegraphics[width=0.43\textwidth]{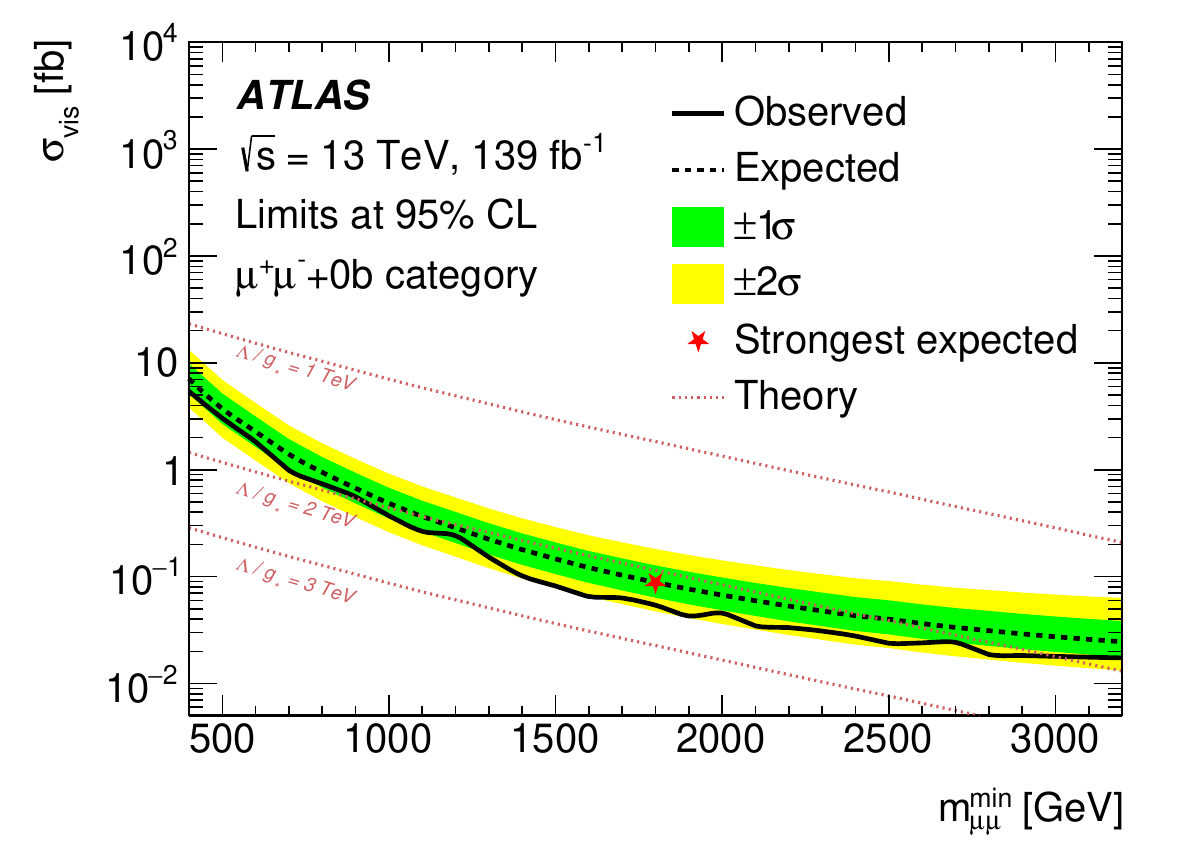}
    \includegraphics[width=0.45\textwidth]{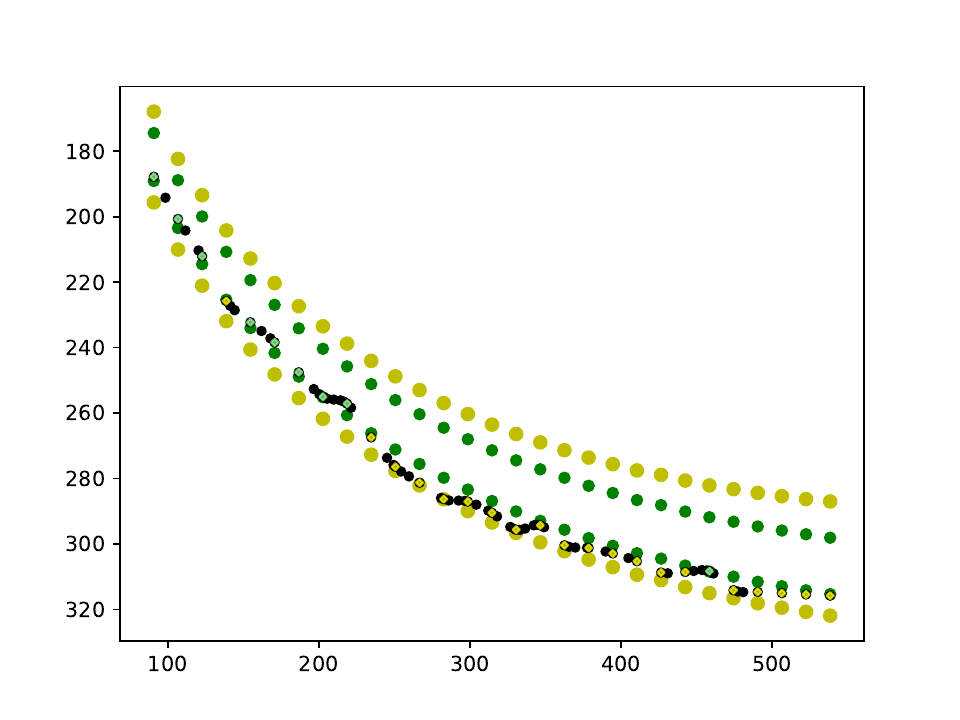}
    \caption{Sample exclusion plot from arXiv:2105.13847~\cite{atlas_2105_13847} (Fig.~4c), left, and what the column scan reads from it, right: the vertices of the boundaries of the $1\sigma$ band (green) and the $2\sigma$ band (yellow), the vertices of the observed limit (black), and, as small diamonds, the points classified.}
    \label{sceleton}
\end{figure*}

The previous algorithm was optimized to analyze standard exclusion plots (e.g., Fig.~\ref{sceleton}). As a result, certain plots, such as those shown in Fig.~\ref{unregularplots}, could not be analyzed due to their atypical structural features. To address these limitations, we developed extensions to the original algorithm to enable the analysis of these non-standard plot structures.

The distinctive characteristics of these plots are readily identifiable. One class of non-standard plots is binned (Fig.~\ref{unregularplots}, left): the $x$-axis is divided into discrete bins, such as ranges of a mass threshold or types of a coupling. Each bin holds a single result, which both allows and requires the extraction of only one representative data point per bin in order to avoid overcounting. Such plots are drawn in several ways. The bands may be separate rectangles, one per bin, or the corners of a single path, and the observed limit may be drawn per bin or as one continuous curve. In each case, the observed limit is reduced to one point per bin and compared with the bands of that bin.

A second class of plots is characterized by uncertainty bands that loop back onto themselves. In these cases, the method described in Section~\ref{sec:method} is not directly applicable as it relies on the natural expectation that a single observed limit corresponds to each value of the $x$-axis parameter. To address this issue, an extension of the algorithm has been implemented to correctly classify and process these looping structures (Section~\ref{sec:classification}).

Two further categories concern how plots are drawn rather than what they show. In the first, the observed limit is drawn as a sequence of horizontal steps, one per bin, so that every analyzed point appears as a pair of vertices at the same height. The polygon method (Section~\ref{sec:polygon}) always merges such pairs into a single point at their midpoint; the column scan does so when this yields more usable points. A bin drawn as several consecutive steps of equal height with the same band edges counts once. In the second, the bands are painted as nested filled regions, one over the other: first the $2\sigma$ region, then the $1\sigma$ region on top of it, and finally a white or differently colored region covering the inner part (e.g., arXiv:1710.01123 and the closed contours of arXiv:2204.11988). What remains visible are the bands, which we reconstruct from the drawing order before the classification. Excluding the looping and binned plots changes the results by at most about one percentage point (Section~\ref{sec:cutoffs}).

\subsection{Data Availability in HEPData}\label{sec:hepdata}
% numbers from data/hepdata/{atlas,cms}-hepdata-check-2026.txt (manual checks; a plot counts as
% available only if the expected and observed limits and both bands are given)
Although the number of HEPData entries has increased over time, the overall data coverage remains limited for the purposes of this analysis. Of the 55 publications of the present collection, 42 have a HEPData record (21 of 28 for ATLAS and 21 of 27 for CMS). We counted a plot as available if its record lists an entry for it with the expected and observed limits and both uncertainty bands. These counts rest on the entries the records list; they are upper limits on the plots that can actually be used, since a closer look at nine publications (see below) showed listed entries that lack one of the bands or whose numbers do not reproduce the drawn curves.

For ATLAS, 114 of the 174 plots of our samples (66\%) are available in this sense. For 31 plots, the publication has a HEPData record that does not include them, and 29 plots belong to publications without a record.

For CMS, 117 of the 219 plots (53\%) are available. A further 55 plots, from three publications, have records whose entries cannot be matched one by one to the figures of the publication; counting them as available would give 79\%. For 25 plots the publication's record does not include them, and 22 plots belong to publications without a record. Of the 117 available plots, 46 are labeled differently on HEPData than in the publication, for example without the letters of the sub-figures or with different figure numbers.

Where HEPData provides the numbers, they also allow a direct check of our reading of the plots. We compared our reading with HEPData for nine publications with records, four from ATLAS and five from CMS (arXiv:2006.05872, 2010.02098, 2304.09553, 2305.15962, 2212.12604, 2305.07998, 2307.08708, 2308.07826 and 2405.09320). For each of their 94 plots in our samples, the table belonging to the plot was identified from the drawn observed curve itself, every tabulated point was classified from the numbers, and the result was compared point by point with the column scan. For 77 plots, a table with both bands reproduces the drawing; in these, 1522 of the 1523 points classified both ways agree, and the one exception is an observed limit that coincides with the edge of the $1\sigma$ band to five significant digits. Of the other 17 plots: (i)~9 have no table that reproduces the drawn observed limit, some of them being missing from the records; (ii)~3 are tabulated with the $1\sigma$ band only; and (iii)~for 5 the tabulated numbers are inconsistent: band offsets with impossible signs or $2\sigma$ offsets equal to the $1\sigma$ ones (4), or an observed limit at a mass for which no expected limit is given (1). All 17 were counted as available above.

Although 231 of these 393 plots (59\%) are available in HEPData, inconsistencies in tabulation structure and presentation limit their direct usability.
For example, in the case of arXiv:2407.10717, the plots available in HEPData differ from those presented on the collaboration’s public results page. In arXiv:2503.20023, while numerical data are provided, the corresponding figures are listed but not displayed. In arXiv:2503.16213, the data corresponding to individual curves are distributed across separate entries, complicating systematic reading and analysis.
More generally, discrepancies in plot labeling between HEPData and the downloaded public results frequently require manual cross-checks for each plot, preventing a fully automated extraction procedure.

\section{Method}\label{sec:method}
All exclusion plots are first downloaded in PDF format from the public results pages of the respective experiments and converted to SVG format with the open-source program \textsc{pdf2svg} (version 0.2.3)~\cite{pdf2svg}. In an SVG file every visual element is a ``path'': a sequence of line and curve segments with attributes such as its fill and stroke colors, its line style and a transformation matrix. We read the files with the Python library \textsc{svgpathtools}~\cite{svgpathtools}, apply each path's transformation, and work with the resulting coordinates, which are lengths on the printed figure in points of 1/72~inch. The analysis then proceeds in three steps: the uncertainty bands and the observed limit are identified among the paths, the observed points that can be used are selected, and each of them is classified by comparing it with the band boundaries in its column.

\subsection{Reading a Plot}\label{sec:reading}
Of all graphical elements of a plot, only the boundaries of the uncertainty bands and the observed limit are used; axes, legends, text, theory predictions and all other curves are ignored (Fig.~\ref{rejectedplots}b,c).

\textit{Uncertainty bands.} The paths of the $1\sigma$ and $2\sigma$ bands are recognized by their colors, using a table of the green and yellow shades found in the collected plots. A plot without a green band is read with the alternative scheme of golden-yellow and sky-blue bands. Of the paths of each color, the two with the most segments are taken and joined into the boundary of that band, unless the longer one has more than three times as many segments, in which case it is used alone. Where the bands are painted as nested regions drawn over each other, the visible parts are reconstructed from the drawing order, each region covering what was drawn before it.

\textit{Observed limit.} The observed limit is drawn as a solid line, in most plots black. Among the solid lines spanning more than a tenth of the figure's width, we take the one with the most segments, in the first color present out of black, brown, dark blue, red and grey. Dashed and dotted lines are not considered, as they draw expected limits and theory predictions. Its vertices are the candidate data points. Paths made of many separate pieces need four rules. (i)~A path of short strokes yields one point per stroke. (ii)~Strokes with a median length above two units are ticks or dashes, and never the observed limit. (iii)~Dots marking analyzed points are rejected when most of them lie on a dashed line, nearer to it than to any solid line; they then mark the expected limit. (iv)~A path consisting of small closed shapes of similar size (markers) gives one point at the center of each. Finally, when several colors offer a candidate, a line whose color has a filled band of nearly the same hue in the plot is taken to be a theory prediction with its uncertainty band, and a candidate of another color is preferred. In a few plots the longest line is not the observed limit. The expected limit may be drawn as a dashed line over a thin solid copy of itself, which then outnumbers the observed line in vertices; or the observed limit may be drawn as dozens of short segments, surrounded by its theory-uncertainty band drawn as two long edge lines. Simple geometric rules recognize these situations and take the observed limit accordingly: by the markers that lie on the observed line, by a dashed curve lying along a solid one, and by short segments that join into a curve between two lines. In our samples they apply to nine plots.

\textit{Visible points.} Observed points that lie outside the clipping region of their curve, or under an opaque white or grey area drawn on top of it, are not visible in the figure and are discarded. Where the observed limit is drawn as steps, the vertex pairs of each step are merged and each bin is counted once (Section~\ref{sec:newalgos}).

\subsection{Classification}\label{sec:classification}
Each remaining observed point is matched to two green and two yellow boundary points at the same $x$-coordinate, which define the lower and upper edges of the $1\sigma$ and $2\sigma$ bands in that column. The $y$-coordinate of the point is then compared with these boundaries to determine whether it lies within the $1\sigma$ band, within the $2\sigma$ band only, or outside both. In looping plots, where the column crosses the bands twice, four boundary points of each color are used.

\begin{figure}[tb]
\centering
\includegraphics[width=0.45\textwidth]{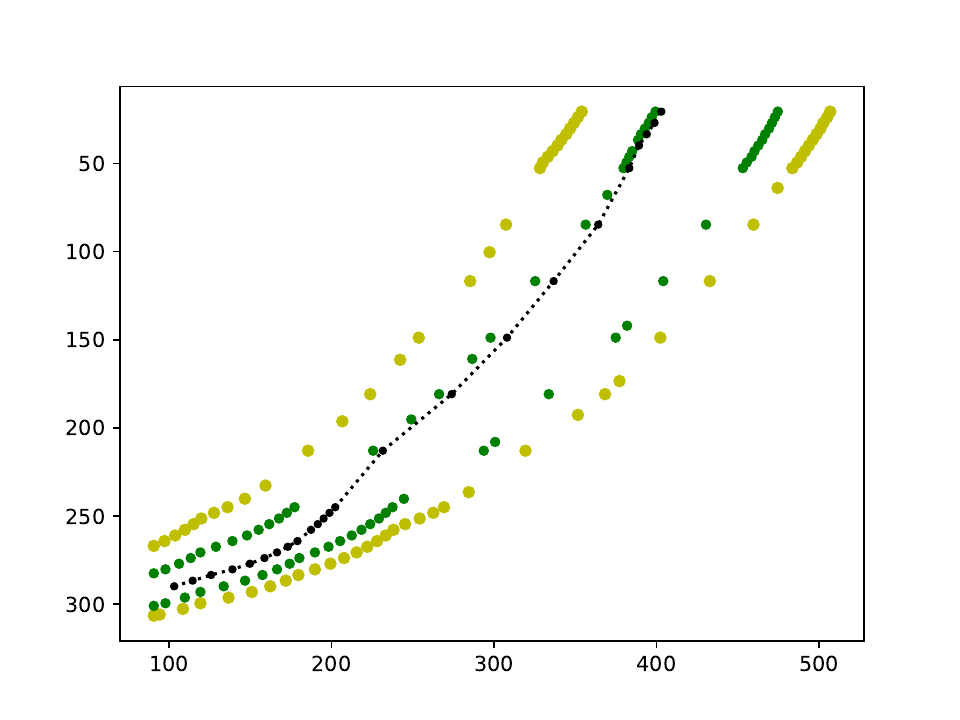}
\caption{The extracted ``skeleton'' of the relevant graphical elements from Fig.~6b of arXiv:2010.02098~\cite{atlas_2010_02098}, for which no data points are retained in the analysis.}
\label{nogoodpts}
\end{figure}

Only points for which such a column can be found are retained, which selects the benchmark points (Section~\ref{sec:data}). If no point of a plot can be matched, the plot is not used (Fig.~\ref{nogoodpts}).

In each column, the column scan expects two boundary points per band: its upper and its lower edge at that $x$. Four features of the drawings can make the matched points something else: (i)~an outline may have two vertices close together on one edge, so that both matched points lie on the same edge; (ii)~where a band is drawn in separate pieces, for instance where the $2\sigma$ band is depicted as two regions above and below the $1\sigma$ band, the two matched points may belong to different pieces and enclose the gap between them rather than the band; (iii)~in such a drawing, the inner edge of a piece may also stand in for its outer edge, if the latter has no vertex in the column; and (iv)~near the crossing of a loop, boundary points of the other branch may fall within the tolerance of a column matched on one branch, at heights that belong to another stretch of the band. The column check recognizes all four by comparing the matched points with the band regions, reconstructed independently from the filled areas of the plot, and rejects the column.

\subsection{Thresholds for the Selection of Reliable Data Points}\label{sec:thresholds}
To ensure reliable classification of data points, we introduce a tolerance for matching the observed points to the band boundaries, and we treat columns in which band boundaries nearly coincide separately. The impact of these choices on the final results is evaluated in Section~\ref{sec:cutoffs}.

\subsubsection{Tolerance Criterion for Reliable Data Point Matching}

Ideally, each black data point is associated with two green and two yellow boundary points at the same $x$-coordinate, allowing a strict classification. In practice, however, the SVG coordinates do not always match exactly and typically differ by a small amount. To prevent such small mismatches from unnecessarily reducing the number of usable data points, a tolerance is introduced: if no boundary points lie at exactly the same $x$-coordinate, those within the tolerance on either side are used, provided that there are exactly two of each color. In looping plots, where the column crosses the bands twice, four boundary points of each color must match exactly. A point is used only if no other observed vertex lies in its column (two, at the same $x$, in a loop) and the $1\sigma$ boundaries lie within the $2\sigma$ boundaries. Larger tolerances increase data coverage but also introduce misclassification by associating unrelated boundary points, while smaller ones reject more points.

SVG coordinates are lengths on the printed figure, in typographic points of 1/72~inch, which we call SVG units. Most figures are 567~SVG units (20~cm) wide, with a plotting frame of about 452~units, but some are drawn considerably smaller or larger; in our collection the width of the plotting frame varies by a factor of almost three. We therefore express the tolerance, like every other distance threshold of the analysis, relative to the width of the plotting frame, taken as the length of its horizontal axis. The tolerance is $\pm 1/452$ of the frame width, i.e.\ $\pm 1$~SVG unit for a frame of standard size. The 2019 study used a fixed tolerance of $\pm 1$~SVG unit for every plot; this choice is treated as a systematic variation in Section~\ref{sec:cutoffs}.

\subsubsection{Columns with Nearly Coinciding Band Boundaries}\label{sec:narrow}
\begin{figure*}[tb]
    \centering
    \includegraphics[width=0.24\textwidth]{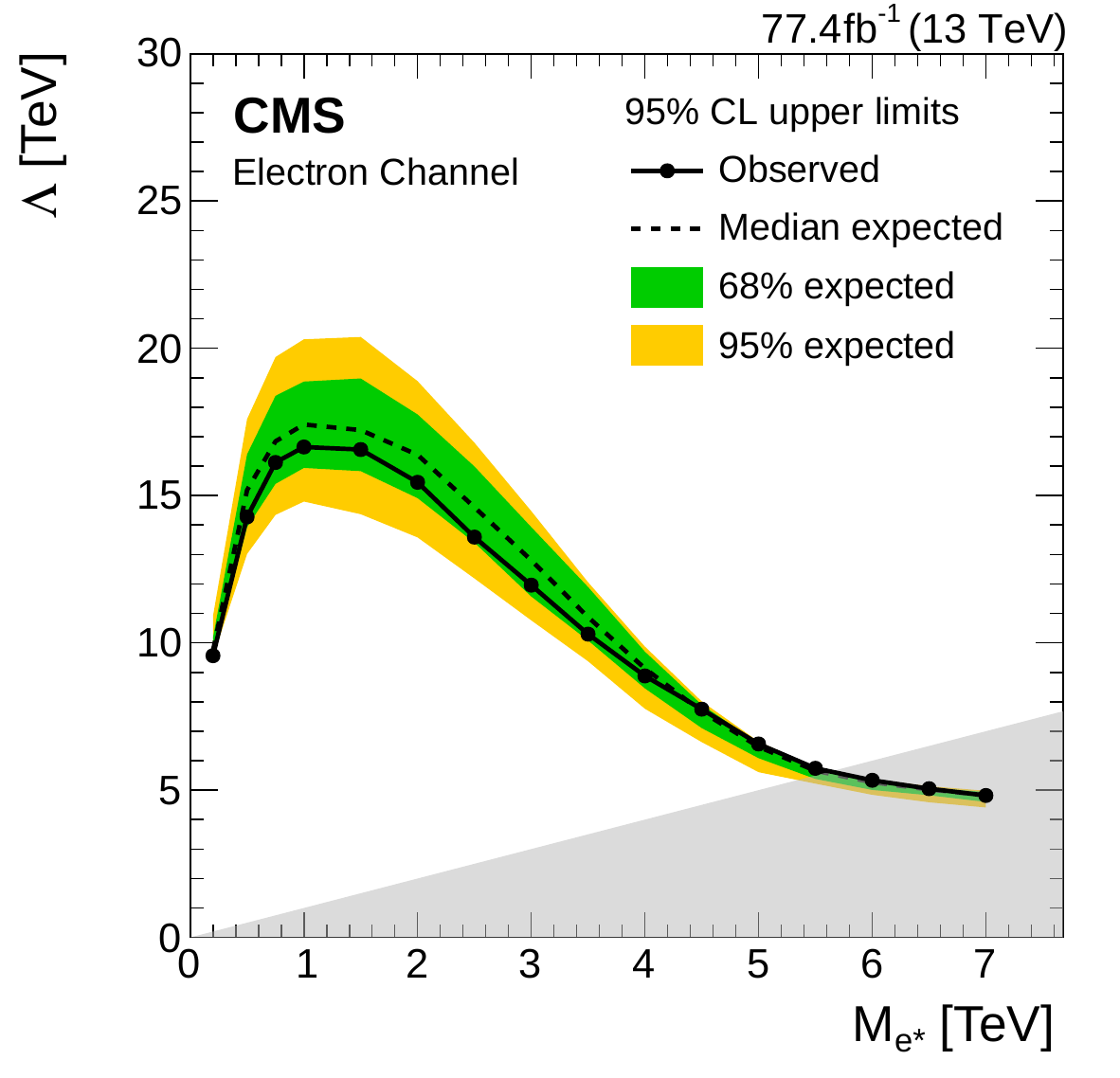}
    \includegraphics[width=0.33\textwidth]{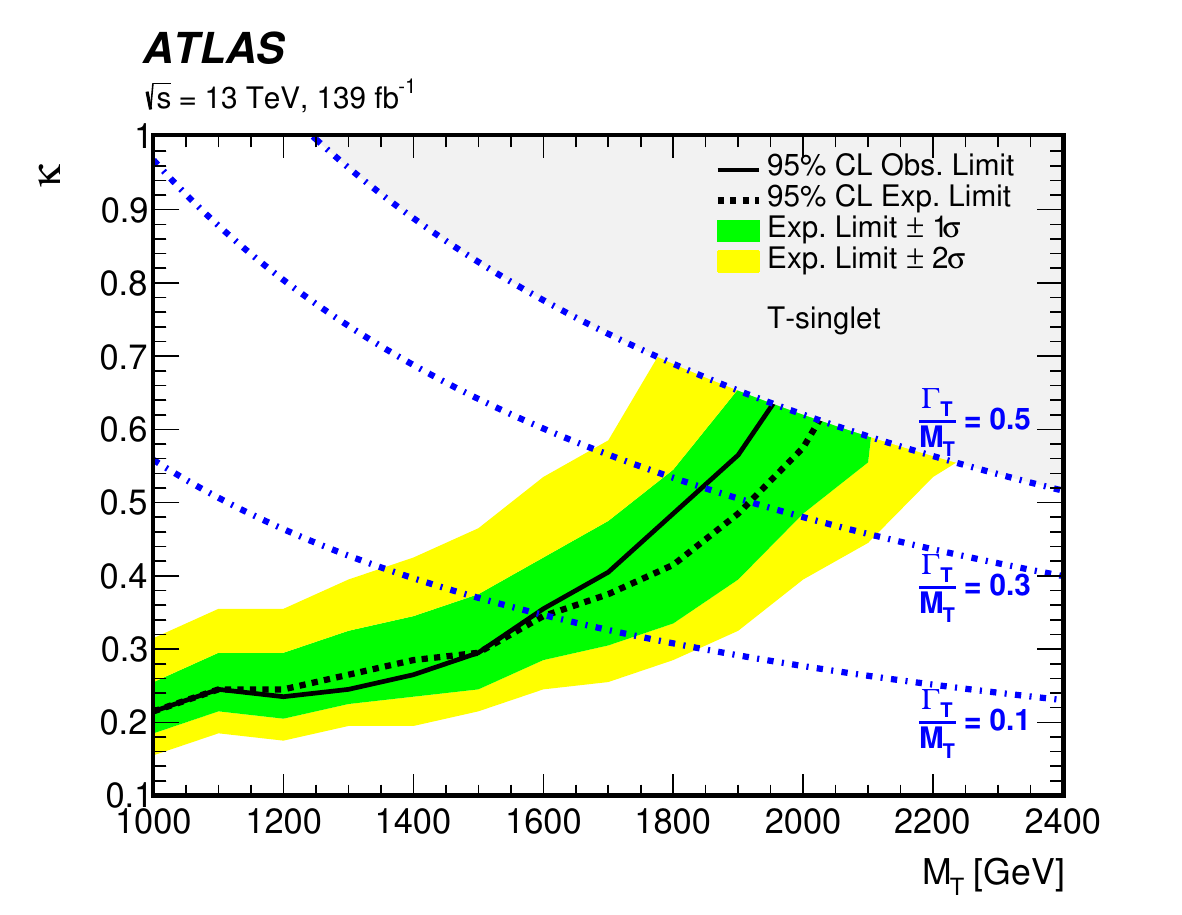}
    \includegraphics[width=0.33\textwidth]{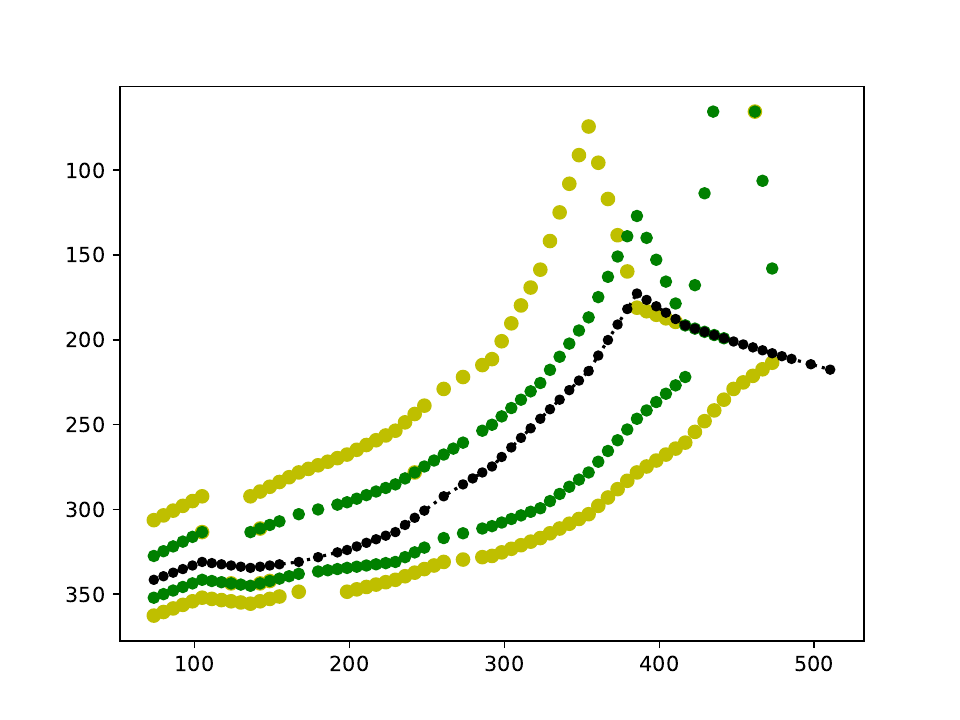}
    \caption{(a) An exclusion plot from arXiv:2001.04521~\cite{cms_2001_04521} (Fig.~10a), in which the uncertainty bands lie in very close proximity. (b,c) An exclusion plot from arXiv:2307.07584~\cite{atlas_2307_07584} (Fig.~9a) and its corresponding ``skeleton'' representation, showing graphical paths excluded from the analysis as structurally irrelevant.}
    \label{rejectedplots}
\end{figure*}

In some columns two of the four band boundaries lie almost on top of each other (e.g., Fig.~\ref{rejectedplots}a). Mostly the $-1\sigma$ and $-2\sigma$ boundaries merge, as happens where the expected background is small; occasionally the merging comes from how the plot was drawn. The 2019 study discarded the data points of such columns, using a threshold on the relative distance between two boundaries. A point near two merged boundaries is no harder to classify than a point near any single boundary, however, and discarding those columns removes a whole regime from the analysis, whatever the observed limit does there. We therefore keep all such points, and make the regime explicit instead.

For every classified point we characterize the bands in its column, independently of the observed limit, by
\begin{equation}
m = \frac{\min(\Delta_\mathrm{low}, \Delta_\mathrm{up})}{w_{1\sigma}},
\end{equation}
where $\Delta_\mathrm{low}$ and $\Delta_\mathrm{up}$ are the distances between the $1\sigma$ and $2\sigma$ boundaries below and above the $1\sigma$ band, and $w_{1\sigma}$ is the width of the $1\sigma$ band. For Gaussian-like bands $m$ is close to 0.5; it tends to zero where boundaries merge. Columns with $m<0.2$ form a \emph{narrow-band} batch, the others a \emph{wide-band} batch; the boundary was chosen so that the narrow batch holds a number of points of the same order as the 2019 threshold discarded. The results of the two batches are compared in Section~\ref{sec:batches}, and discarding the narrow-boundary columns as in 2019 is treated as a variation in Section~\ref{sec:cutoffs}.

\subsection{Averaging Procedures}\label{sec:averaging}\label{sec:dedup}
After counting, for each plot, the number of data points falling within the $1\sigma$ and $2\sigma$ regions as well as those lying outside both, we combine these counts using two different averaging techniques, each leading to a distinct interpretation of the results. The first method, which we call raw averaging, gives equal weight to every data point, whereas the second method, termed normalized averaging, considers each plot as a single data point. This approach prevents the possible dominance of plots with a relatively large number of data points from disproportionately influencing the combined result, particularly when their internal distributions deviate from the overall trend.

It should be emphasized that, rather than applying the averaging procedures independently to all plots, we perform the averaging over groups of figures that are expected to be highly correlated. Exact duplicates, such as a figure that reappears as an auxiliary figure, are removed beforehand by comparing a geometric fingerprint of each plot, consisting of its classified data points and band boundaries in coordinates normalized to the plot extent. In the present collection, near copies are removed as well: auxiliary or appendix figures that repeat parts of another plot of their publication, redrawn or over another range. They are recognized by the position of each classified point relative to the bands of its column, which does not change when a plot is redrawn; two plots of one publication that share a run of consecutive points with the same positions, at least half of the smaller plot's points but no more than 30 and no fewer than 3, show the same result, and the auxiliary or appendix one is removed. This removes nine plots. The publications of the 2019 study keep their original selection (Section~\ref{sec:revisit}). In several publications, the collaborations present multiple exclusion plots of one figure, its subfigures, which typically show the same search under varied assumptions. Plots of one publication whose file names differ only in the subfigure letter (or, for unnumbered figures, in a trailing number) form a group. To avoid biasing the meta-analysis through repeated use of highly similar information, we weight the data points extracted from such groups so that they are treated collectively as originating from a single plot. For convenience, these groups of highly similar plots are hereafter referred to as a single plot throughout the analysis.

\subsubsection*{Raw Averaging}
Initially, counts from all plots are collected, and counts from highly similar plots are averaged to obtain a single representative value for each group, as described above. These representative counts are then summed over all groups, and the corresponding percentages are computed as simple ratios. We call the mean number of classified points over the plots of a plot group the \emph{size} of the group. The size of a set of plot groups, such as a publication or a whole collection, is the sum of their sizes. The raw fraction of a set is thus the sum of the representative counts inside a band divided by its size. The evaluation of the statistical uncertainty is described in Section~\ref{sec:bootstrap}.

\subsubsection*{Normalized Averaging}
In this second approach, each plot group is treated as a single effective data point, independent of how many measurements it contains. This is achieved by weighting the data points so that each plot group contributes one in total. A point's weight is the reciprocal of the product of the number of plots in its group and the number of data points of its plot.

Using these effective counts, we compute the corresponding percentages for each plot through simple ratios. The resulting plot-level fractions are then combined by averaging to obtain the overall fractions. The evaluation of the statistical uncertainty is described in Section~\ref{sec:bootstrap}.

\subsection{Statistical Uncertainties}
\label{sec:bootstrap}

A simple counting error does not describe the uncertainty of the fractions defined above, for three reasons. First, the data points of a single plot are not independent: neighboring points share the same background model and systematic uncertainties, and the signal mass resolution is often coarser than the spacing of the points. A binomial error computed from the number of points therefore overstates the amount of independent information. Second, the different plot groups of one publication are derived from the same data set and the same analysis, so they are correlated as well. Third, and most importantly, the coverage need not be the same for every analysis. Searches differ in their statistical regime, in their treatment of systematic uncertainties and in the approximations used to compute the expected bands, so the fraction of points inside the bands genuinely varies from one plot to another. This scatter is a source of uncertainty on the average that no counting error can capture. The binomial error assumes that all points are drawn with one common probability, and the error propagated through the normalized averaging only accounts for the counting fluctuations within each plot group.

We therefore estimate the statistical uncertainties with a nonparametric bootstrap~\cite{efron1979,efron1993}, resampling entire publications rather than individual points or plots. The set of $N$ analyzed publications of an experiment is regarded as a sample drawn from the population of searches of that experiment. A bootstrap replica is formed by drawing $N$ publications from this set at random with replacement, so that some publications appear several times and others not at all. The complete analysis chain, including the grouping of similar plots and both averaging procedures, is then repeated on the replica. A publication drawn more than once is entered under a distinct label for each copy, so that the grouping procedure treats the copies as separate publications rather than merging them. Repeating this for $B$ replicas yields fractions $\hat f^{*}_{1},\dots,\hat f^{*}_{B}$, and the statistical uncertainty of the nominal fraction $\hat f$ is taken as their standard deviation,
\begin{equation}
\begin{split}
\sigma_{\rm boot}^{2} &= \frac{1}{B}\sum_{b=1}^{B}\left(\hat f^{*}_{b}-\bar f^{*}\right)^{2},\\
\bar f^{*} &= \frac{1}{B}\sum_{b=1}^{B}\hat f^{*}_{b}.
\end{split}
\end{equation}
For fractions close to 100\%, where the bootstrap distribution becomes asymmetric, we also give the central interval containing 68\% of the replicas.

Resampling at the level of publications has the advantage that all correlations within a publication, between its points as well as between its plots, are carried along automatically without having to be modeled. The variation between publications enters directly, since the replicas differ precisely in which publications they contain. The publication is the largest unit that can be considered approximately independent; possible correlations between different publications, for instance between analyses of overlapping data sets, are not covered by this procedure and are discussed separately in Section~\ref{sec:validation}. Since the number of publications per experiment is modest (\nPapersCms{} for CMS and \nPapersAtlas{} for ATLAS), the bootstrap uncertainties are themselves only approximate. For $N$ independent units, the relative standard error of an estimated standard deviation is roughly $1/\sqrt{2(N-1)}$, about 14\% here, and inference based on a small number of resampled clusters is known to be approximate in general~\cite{cameron2008}. This precision is adequate for the conclusions drawn here.

% factors from the analyzer's binomial/propagated errors (*.ana) against the bootstrap (*.boot),
% column scan, current and 2019 collections (git 3e3e821): 2.1--8.8
The bootstrap uncertainties are larger than the naive estimates by factors of three to nine for the $1\sigma$ fractions and one to five for the $2\sigma$ fractions, the largest increases occurring for normalized averaging, where the scatter between plot groups dominates. Unless stated otherwise, all uncertainties quoted in this paper are bootstrap uncertainties.

As a complementary diagnostic, we repeat the analysis leaving out one publication at a time. This identifies results that are dominated by a single publication, which is relevant here because the number of extracted points varies by more than two orders of magnitude between publications.

\section{Results and Discussion}\label{sec:results}
The results of the analysis are given in Table~\ref{tab:results}, with bootstrap uncertainties (Section~\ref{sec:bootstrap}). For nominally calibrated bands, 68.3\% of the points would lie within the $1\sigma$ band and 95.4\% within the $2\sigma$ band. We regard normalized averaging, in which no plot group outweighs another, as the primary statistic, and raw averaging, which continues the 2019 analysis, as the secondary one.

\begin{table*}[tb]
\centering
\topcaption{Fractions of points within the expected-limit bands for the present collection (column scan, bootstrap uncertainties).}\label{tab:results}
% generated by analysis/papertables.py from base git 2c7a2e8, 2026-11-03; do not edit
\begin{tabular}{lcccc}
\toprule
 & \multicolumn{2}{c}{CMS} & \multicolumn{2}{c}{ATLAS} \\
\cmidrule(lr){2-3}\cmidrule(lr){4-5}
 & Raw & Normalized & Raw & Normalized \\
\midrule
Size (raw), plot groups (normalized) & 1886.9 & 104 & 1104.2 & 64 \\
Fraction in $1\sigma$ band (\%) & $69.5 \pm 3.2$ & $66.1 \pm 4.4$ & $81.8 \pm 5.9$ & $73.3 \pm 7.6$ \\
Fraction in $2\sigma$ band (\%) & $95.8 \pm 1.2$ & $95.3 \pm 1.9$ & $99.5 \pm 0.3$ & $99.2 \pm 0.4$ \\
\bottomrule
\end{tabular}

\end{table*}

For CMS, the results agree with the nominal coverage: $\resRawOneCms\%$ of the points lie within the $1\sigma$ band and $\resRawTwoCms\%$ within the $2\sigma$ band for raw averaging, and $\resNormOneCms\%$ and $\resNormTwoCms\%$ for normalized averaging, all within about one standard deviation of the nominal values.

For ATLAS, the fraction within the $1\sigma$ band is $\resRawOneAtlas\%$ for raw averaging, \devRawOneAtlas{} standard deviations above the nominal value, and $\resNormOneAtlas\%$ for normalized averaging, \devNormOneAtlas{} standard deviations above it. The $2\sigma$ fractions, $\resRawTwoAtlas\%$ and $\resNormTwoAtlas\%$, lie clearly above the nominal 95.4\%: very few observed limits fall outside the $2\sigma$ band. Taken together, the ATLAS expected-limit bands contain the observed limit more often than their nominal probability; the evidence is strongest for the $2\sigma$ band. Possible causes are discussed in Section~\ref{sec:interpretation}.

The difference between raw and normalized averaging reflects how the plots are weighted. Raw averaging weights each plot group by its size, so that publications with many points weigh more; normalized averaging gives every plot group the same weight. Leaving out one publication at a time shows how far single publications drive the results. For ATLAS, the raw $1\sigma$ fraction changes by at most $+\looUpRawOneAtlas$ and $-\looDownRawOneAtlas$ percentage points (without \looUpPubRawOneAtlas{} and \looDownPubRawOneAtlas{}, respectively), and the normalized one by at most $+\looUpNormOneAtlas$ points (without \looUpPubNormOneAtlas{}). For CMS, the corresponding changes are at most $+\looUpRawOneCms$ and $-\looDownRawOneCms$ points for raw and $+\looUpNormOneCms$ and $-\looDownNormOneCms$ points for normalized averaging. No single publication determines the conclusions, but individual publications move the ATLAS results by up to about one bootstrap standard deviation, consistent with the size of these uncertainties.

Compared with the 2019 study (Table~\ref{tab:revisit2019}), the CMS $1\sigma$ fraction has moved closer to the nominal value, while the ATLAS fractions have moved above it. Whether the ATLAS excess is a feature of the low-background regime, where the $-1\sigma$ and $-2\sigma$ boundaries merge, is examined with the narrow- and wide-band batches in Section~\ref{sec:batches}: the raw $1\sigma$ fraction of the wide-band batch remains above the nominal value. For CMS, too, the points in the narrow-band batch lie more often within the $1\sigma$ band than those in the wide-band batch, but the wide-band batch agrees with the nominal coverage.

\subsection*{Narrow and Wide Bands}\label{sec:batches}
Table~\ref{tab:batches} compares the coverage of the narrow-band and wide-band batches defined in Section~\ref{sec:narrow}. The narrow batch holds only 3--7\% of the points, so its uncertainties are large. In the 2019 collection the two batches agree within their uncertainties. In the present collection, the narrow batch shows a larger fraction of points within the $1\sigma$ band in both experiments. With the batch boundary at $m<0.1$ or $m<0.3$ instead of $m<0.2$, the normalized fractions of the wide batch change by less than two percentage points, while the narrow-band results move within their uncertainties.

\begin{table*}[tb]
\centering
\topcaption{Fractions of points within the bands (\%) for the narrow-band ($m<0.2$) and wide-band batches, column scan, with bootstrap uncertainties.}\label{tab:batches}
% generated by analysis/papertables.py from base git 2c7a2e8, 2026-11-03; do not edit
\small
\setlength{\tabcolsep}{3.5pt}
\begin{tabular}{llccccc}
\toprule
 & Bands & Points & Raw $1\sigma$ & Raw $2\sigma$ & Norm. $1\sigma$ & Norm. $2\sigma$ \\
\midrule
ATLAS & narrow & 186 & $95.6 \pm 5.4$ & $99.3 \pm 0.7$ & $89.6 \pm 5.6$ & $98.3 \pm 1.1$ \\
 & wide & 2305 & $80.0 \pm 5.9$ & $99.6 \pm 0.3$ & $73.3 \pm 7.7$ & $99.5 \pm 0.3$ \\
CMS & narrow & 230 & $82.7 \pm 2.9$ & $100.0 \pm 0.0$ & $85.0 \pm 3.0$ & $100.0 \pm 0.0$ \\
 & wide & 4780 & $68.8 \pm 3.4$ & $95.7 \pm 1.2$ & $65.8 \pm 4.7$ & $95.3 \pm 1.9$ \\
ATLAS 2019 & narrow & 228 & $70.0 \pm 6.2$ & $97.6 \pm 1.4$ & $67.2 \pm 6.2$ & $96.5 \pm 2.4$ \\
 & wide & 7975 & $69.4 \pm 2.2$ & $96.5 \pm 1.2$ & $65.1 \pm 3.6$ & $94.4 \pm 1.8$ \\
CMS 2019 & narrow & 449 & $77.9 \pm 7.3$ & $96.3 \pm 2.8$ & $68.8 \pm 6.8$ & $90.7 \pm 6.0$ \\
 & wide & 7085 & $74.9 \pm 1.7$ & $97.0 \pm 0.9$ & $76.4 \pm 3.3$ & $97.5 \pm 0.7$ \\
\bottomrule
\end{tabular}

\end{table*}

\subsection*{Interpretation}\label{sec:interpretation}
A share of observed limits within the bands above the nominal one does not by itself mean overestimated uncertainties. Three mechanisms can produce it: uncertainties estimated conservatively, which widen the bands; bands computed a posteriori, from pseudo-data built with the background-only fit to the observed data, which follow the observed fluctuation; and the low-count regime, where the number of events is discrete and the asymptotic formulae fail. A-posteriori bands are the default of the common statistical tools and, according to a discussion between reinterpretation groups, tool developers and members of both collaborations~\cite{pyhf_discussion1619}, of the published ATLAS and CMS limits; the same discussion shows that they surprise even experienced users, who read ``expected'' as ``before the observation''. The publications do not say which convention they used. To judge the size of these effects, signal-free pseudo-experiments have been conducted in simple counting models with asymptotic CL$_s$ limits, computed with \textsc{pyhf}~\cite{pyhf_joss,pyhf_zenodo}. Based on their results, we make the following observations. (i)~Without background uncertainty, the fractions are close to but slightly above the nominal values, because the observed counts are integers: for 50 expected background events, 71\% and 97\% of the observed limits lie within the $1\sigma$ and $2\sigma$ bands. (ii)~Where a control region fixes the background normalization, a-priori and a-posteriori bands give the same fractions. (iii)~A-posteriori bands raise the fractions substantially only where a large uncertainty, half the background or more, is constrained by the signal region itself; then up to about 99\% of the observed limits lie within the $2\sigma$ band. (iv)~Few background events raise the fractions whatever the convention: with one expected event, 92\% lie within the $1\sigma$ band and none below it, since CL$_s$ limits cannot become much stronger than expected when hardly any events are observed.

The mechanisms differ in which side of the band the observed limits avoid: few events cannot give a limit much stronger than expected and empty mainly the lower side, while bands that are too wide leave both sides short. For ATLAS, the points are short below the $1\sigma$ band, as the low-count regime predicts, but almost none lie beyond either $2\sigma$ edge, which few events alone do not produce. The ATLAS results are therefore best explained by bands that are wider than the spread of the observed limits, together with the low-count regime. Such bands can result from conservatively estimated uncertainties, or from expected limits computed a posteriori where a signal region constrains a large uncertainty itself. Given that most analyses fit the level of their backgrounds with control regions, a-posteriori computation is unlikely to account for the whole effect.

\section{Validation and Consistency Checks}\label{sec:validation}
The qualitative change observed between the results of the previous study and the present analysis\,---\allowbreak\,where CMS now appears more consistent while ATLAS exhibits a noteworthy deviation\,---\allowbreak\,necessitated a careful and thorough reassessment of our methodology. Consequently, a series of validation and consistency checks were performed to ensure the robustness of both the updated code and the analysis strategy.

The checks in this section concern the comparison with the 2019 study and the consistency between publication periods (Sections~\ref{sec:revisit} and~\ref{sec:time}), the reading of the plots (Sections~\ref{sec:polygon}--\ref{sec:markers}), the statistical treatment (Sections~\ref{sec:separated} and~\ref{sec:independence}), and the choices made in the analysis (Section~\ref{sec:cutoffs}).

\subsection{Revisiting and Extending the 2019 Dataset}\label{sec:revisit}
\begin{table}[tb]
\centering
\topcaption{arXiv identifiers of the publications of the 2019 collection that did not enter the 2019 analysis. The publications of the original analysis are listed in Ref.~\cite{aciksoz2019}.}\label{tab:papers2019}
% generated by analysis/papertables.py from base git 2c7a2e8, 2026-11-03; do not edit
\begin{tabular}{cc@{\hspace{2em}}c@{\hspace{2em}}cc}
\toprule
\multicolumn{2}{c}{ATLAS} & & \multicolumn{2}{c}{CMS} \\
\midrule
1801.06992 & 1706.03948 &  & 1806.03472 & 1803.11116 \\
1710.01123 & 1609.04572 &  & 1806.01058 & 1803.06553 \\
1709.10440 & 1607.05621 &  & 1806.00408 & 1802.09407 \\
1709.07242 & 1606.04782 &  & 1805.10191 & 1802.01486 \\
1707.06958 &  &  & 1805.06013 & 1711.04652 \\
1707.03347 &  &  & 1805.04865 & 1711.04370 \\
\bottomrule
\end{tabular}

\end{table}
\label{sec:revisit2019}
As a first validation, the updated analysis code was applied to the exclusion plots collected for the 2019 study~\cite{aciksoz2019}. This collection is larger than the set of publications that entered the 2019 results: it comprises 32 ATLAS and 46 CMS publications from 2016--2018, including all 22 ATLAS and 34 CMS publications of the original analysis; the additional ones are listed in Table~\ref{tab:papers2019}. One of the latter, arXiv:1706.04260, contributes no usable plot with the present tolerance (Section~\ref{sec:cutoffs}), so that 33 and 45 CMS publications enter Table~\ref{tab:revisit2019}. We give results both for the publications of the original analysis, which allow a direct comparison with the published values, and for the full collection, the most complete sample of that period available to us. Duplicate plots are removed automatically (Section~\ref{sec:dedup}); in this collection three ATLAS plots were identified as duplicates, two of them auxiliary copies of a figure in the same publication and one a figure that appears in two different publications.

The results are shown in Table~\ref{tab:revisit2019}. For the publications of the original analysis, all fractions agree with the published values within about one bootstrap standard deviation. The ATLAS publications of the original analysis include six auxiliary figures of arXiv:1805.01908 that repeat parts of its Figs.~8a--c and make up about 38\% of the size of that collection; the 2019 study counted them as well, so they are kept here for the comparison. Leaving them out lowers the ATLAS fractions by at most 1.3 percentage points. For the full collection, the largest difference with respect to the published values, in the CMS $2\sigma$ fraction, amounts to about two bootstrap standard deviations. Judged by the binomial uncertainties of the 2019 study, the same difference would appear highly significant, which illustrates why these uncertainties cannot be used to compare results (Section~\ref{sec:bootstrap}). We conclude that the modifications of the analysis code do not by themselves shift the results, so that differences between the 2019 and the present results have to be sought in the analyzed publications.

\begin{table*}[tb]
\centering
\topcaption{Results of the updated analysis code (raw averaging) for the 2019 dataset, for the publications of the original analysis and for the full collection, compared with the values published in 2019~\cite{aciksoz2019}. The first uncertainty is the binomial one, as in the 2019 study, so that the values can be compared with the published ones; the uncertainty in brackets is the bootstrap uncertainty of Section~\ref{sec:bootstrap} ($B=1000$ replicas), used throughout this paper.}\label{tab:revisit2019}
% generated by analysis/papertables.py from base git 2c7a2e8, 2026-11-03; do not edit
\small
\setlength{\tabcolsep}{4pt}
\begin{tabular}{lcccc}
\toprule
 & \multicolumn{2}{c}{CMS} & \multicolumn{2}{c}{ATLAS} \\
\cmidrule(lr){2-3}\cmidrule(lr){4-5}
 & 2019 publications & Full collection & 2019 publications & Full collection \\
\midrule
Publications & 33 & 45 & 22 & 32 \\
Size & 2454.58 & 3252.08 & 4133.96 & 4625.63 \\
In $1\sigma$ band (\%) & $75.90 \pm 0.86$ [$\pm 2.2$] & $75.04 \pm 0.76$ [$\pm 1.6$] & $69.54 \pm 0.72$ [$\pm 2.6$] & $69.43 \pm 0.68$ [$\pm 2.1$] \\
In $2\sigma$ band (\%) & $96.55 \pm 0.37$ [$\pm 1.0$] & $96.98 \pm 0.30$ [$\pm 0.9$] & $96.38 \pm 0.29$ [$\pm 1.5$] & $96.53 \pm 0.27$ [$\pm 1.1$] \\
\midrule
\multicolumn{5}{l}{\textit{Published in 2019}} \\
Size & 2206.75 & & 4157.33 & \\
In $1\sigma$ band (\%) & $74.04 \pm 0.93$ & & $68.98 \pm 0.72$ & \\
In $2\sigma$ band (\%) & $95.23 \pm 0.45$ & & $96.03 \pm 0.30$ & \\
\bottomrule
\end{tabular}

\end{table*}

For the full collection, the size is larger than in the 2019 study. This has two causes: the collection used here includes the additional publications mentioned above, and the newly implemented algorithm for dotted observed curves recovers plots, and hence data points, that the earlier code could not process and therefore excluded.

\subsection{Consistency Between Publication Periods}\label{sec:time}
As a consistency check, we divided each experiment's present collection into two independent subsets by year of publication, 2020--2022 and 2023--2025, and compared them with each other and with the full collection.\footnote{Although the analyses are not necessarily conducted in the exact year of publication, we use the year indicated in the corresponding arXiv identifier as a reference.} The results are given in Table~\ref{tab:timesplit}; since the two subsets contain different publications, their uncertainties are independent.

For CMS, the earlier subset has $\tsRawOneCmsEarly\%$ of the points within the $1\sigma$ band for raw averaging, and the later subset $\tsRawOneCmsLate\%$; the differences between the two subsets amount to \tsDiffMinCms{} to \tsDiffMaxCms{} standard deviations, depending on the fraction. For ATLAS, the corresponding fractions are $\tsRawOneAtlasEarly\%$ and $\tsRawOneAtlasLate\%$, and the differences amount to at most \tsDiffMaxAtlas{} standard deviations. The subsets of each experiment thus agree with each other within their uncertainties. The ATLAS raw fractions and the $2\sigma$ fractions lie above the nominal values in both periods, while the CMS fractions scatter around them, the normalized ones by up to about two to three standard deviations.

\begin{table*}[tb]
\centering
\topcaption{Results for the publications of 2020--2022 and of 2023--2025, according to the year of the arXiv identifier (column scan, bootstrap uncertainties).}\label{tab:timesplit}
% generated by analysis/papertables.py from base git 2c7a2e8, 2026-11-03; do not edit
\small
\setlength{\tabcolsep}{3.5pt}
\begin{tabular}{llcccccc}
\toprule
 & Years & Publications & Plots & Raw $1\sigma$ & Raw $2\sigma$ & Norm. $1\sigma$ & Norm. $2\sigma$ \\
\midrule
CMS & 2020--2022 & 13 & 118 & $72.7 \pm 3.8$ & $97.5 \pm 1.3$ & $70.4 \pm 5.2$ & $98.1 \pm 1.0$ \\
 & 2023--2025 & 14 & 87 & $65.3 \pm 5.3$ & $93.6 \pm 1.9$ & $57.0 \pm 7.0$ & $89.5 \pm 3.9$ \\
ATLAS & 2020--2022 & 12 & 72 & $87.0 \pm 6.1$ & $99.4 \pm 0.6$ & $82.1 \pm 4.2$ & $98.4 \pm 0.7$ \\
 & 2023--2025 & 16 & 96 & $75.9 \pm 8.6$ & $99.6 \pm 0.2$ & $68.0 \pm 10.8$ & $99.7 \pm 0.2$ \\
\bottomrule
\end{tabular}

\end{table*}

\subsection{Cross-Check with the Polygon Method}\label{sec:polygon}
The column scan classifies a point by comparing it with the band boundaries found in its own column (Section~\ref{sec:classification}). As an independent cross-check, every observed point was also classified by testing whether it lies inside the polygons formed by the visible band regions, computed with the library \textsc{Shapely}~\cite{shapely}, a method that does not rely on matching vertices in a column and therefore accepts more points. Table~\ref{tab:methods} compares the two methods. A point lying on a band boundary, within 0.01 units, is counted half in each of the two classes meeting there; such exact coincidences concern 0.2--1.5\% of the points, depending on the collection. Since both methods read the same plots of the same publications, their results are strongly correlated, and the uncertainty of their difference is obtained from a joint bootstrap, in which each replicate computes both methods on the same sample of publications. In normalized averaging, the methods differ by at most 2.0 percentage points (ATLAS, $1\sigma$ band, present collection), much less than the statistical uncertainties of the results. Because the two methods are strongly correlated, even such small differences reach up to 2.6 standard deviations (ATLAS, 2019 collection, $1\sigma$); in every sample, the polygon method classifies slightly fewer points within the $1\sigma$ band. In raw averaging the differences are larger, up to $\methMaxDiff$ percentage points (\methMaxWhat, \methMaxSig{} standard deviations), since the polygon method also uses the vertices of the observed curve between the analysed values, about twice as many points. The conclusions of Section~\ref{sec:results} are the same for both methods.

\begin{table*}[tb]
\centering
\topcaption{Fractions of points within the bands (\%) obtained with the column scan and with the polygon method, with bootstrap uncertainties.}\label{tab:methods}
% generated by analysis/papertables.py from base git 2c7a2e8, 2026-11-03; do not edit
\small
\setlength{\tabcolsep}{3.5pt}
\begin{tabular}{lcccccccc}
\toprule
 & \multicolumn{2}{c}{Raw $1\sigma$} & \multicolumn{2}{c}{Raw $2\sigma$} & \multicolumn{2}{c}{Norm. $1\sigma$} & \multicolumn{2}{c}{Norm. $2\sigma$} \\
\cmidrule(lr){2-3}\cmidrule(lr){4-5}\cmidrule(lr){6-7}\cmidrule(lr){8-9}
 & column & polygon & column & polygon & column & polygon & column & polygon \\
\midrule
ATLAS 2019 & $69.4 \pm 2.1$ & $71.4 \pm 3.6$ & $96.5 \pm 1.1$ & $96.7 \pm 1.0$ & $65.4 \pm 3.5$ & $64.5 \pm 3.5$ & $94.5 \pm 1.9$ & $94.0 \pm 1.9$ \\
CMS 2019 & $75.0 \pm 1.6$ & $72.3 \pm 1.9$ & $97.0 \pm 0.9$ & $96.8 \pm 0.7$ & $75.1 \pm 3.4$ & $74.9 \pm 3.1$ & $96.2 \pm 1.6$ & $96.7 \pm 1.1$ \\
ATLAS & $81.8 \pm 5.9$ & $84.6 \pm 4.7$ & $99.5 \pm 0.3$ & $97.6 \pm 1.6$ & $73.3 \pm 7.6$ & $71.2 \pm 7.8$ & $99.2 \pm 0.4$ & $98.8 \pm 0.5$ \\
CMS & $69.5 \pm 3.2$ & $73.3 \pm 3.7$ & $95.8 \pm 1.2$ & $96.2 \pm 1.0$ & $66.1 \pm 4.4$ & $65.5 \pm 4.5$ & $95.3 \pm 1.9$ & $95.2 \pm 1.9$ \\
\bottomrule
\end{tabular}

\end{table*}

\subsection{Pixel Check}\label{sec:pixcheck}
The geometric reading of the vector graphics can be verified against the picture that a reader sees. Each plot was rendered with \textsc{CairoSVG}~\cite{cairosvg} with only its band fills visible, and the color of the rendered image at the position of every classified point was compared with the class assigned by the analysis. For the present and the 2019 collections together, all \pixComparedCol{} points of the column scan that could be compared agree with the rendered color, and so do all \pixComparedPoly{} points of the polygon method. Points too close to a band boundary to be decided by the color of the rendered image (\pixEdgeCol{} and \pixEdgePoly{}, respectively) were not compared. This check cannot detect a wrongly identified observed curve, since it only tests the classification of the points it is given.

\subsection{Comparison with the Markers of the Observed Limit}\label{sec:markers}
% numbers from data/legendcheck/markercheck-2026-10.tsv (analysis/markercheck.py, loose match)
In many plots the observed limit is drawn with a marker at each value for which the limit was computed. These markers provide an independent reference for the column scan, which aims at exactly these values rather than at the interpolating line drawn between them. This comparison uses the 290 plots of the present and 2019 collections whose observed limit carries markers drawn as separate shapes. Their 6115 markers within the bands were compared with the 6080 points selected by the column scan; we required only that the horizontal positions agree within the matching tolerance of the column scan. For 97.2\% of the markers the column scan selected a point at the same position, and 2.6\% of the selected points have no marker there. The way the bands are drawn contributes substantially to the remaining 2.8\%. About half of the markers without a selected point (88 of 172) lie at values where one of the band edges has no vertex in the drawing, so that the column scan cannot use that column. Such gaps are common: in a quarter to a third of the plots, a band vertex on one edge has no counterpart at the same value on the other. The others lie in columns rejected by the column check or by the other selection criteria. Where the horizontal positions agree, the vertical positions agree as well: within 0.01 units for 99.5\% of the markers. Since markers drawn as part of the line itself cannot be identified as separate shapes, the second number is an upper bound.

\subsection{Subsample Analysis with Maximally Separated Data Points}\label{sec:separated}
As the LHC collaborations use a single set of simulated background events for different parameter values of a given signal, and since the parameter resolution may exceed the spacing of the points in a plot, nearby data points are likely to be correlated. Assuming that these correlations are smallest between the points furthest apart, we repeated the analysis using only the leftmost and the rightmost classified point of every plot (one point for the few plots with a single classified point). The results, given in Table~\ref{tab:separated}, have larger uncertainties but agree with those of the full analysis, suggesting that correlations between neighboring points are not the source of the observed behavior.

\begin{table*}[tb]
\centering
\topcaption{Results using only the two maximally separated data points of each plot (column scan, raw averaging, bootstrap uncertainties).}\label{tab:separated}
% generated by analysis/papertables.py from base git 2c7a2e8, 2026-11-03; do not edit
\begin{tabular}{lcccc}
\toprule
 & CMS & ATLAS & CMS 2019 & ATLAS 2019 \\
\midrule
Plots & 205 & 168 & 188 & 220 \\
Points & 407 & 336 & 375 & 439 \\
Fraction in $1\sigma$ band (\%) & $62.9 \pm 4.2$ & $74.6 \pm 6.4$ & $77.7 \pm 4.1$ & $69.4 \pm 4.8$ \\
Fraction in $2\sigma$ band (\%) & $96.3 \pm 1.6$ & $98.7 \pm 0.8$ & $95.7 \pm 1.6$ & $95.7 \pm 2.0$ \\
\bottomrule
\end{tabular}

\end{table*}

\subsection{Independence of Points, Plot Groups and Publications}\label{sec:independence}\label{sec:citations}
Plot groups are the panels of a figure that differ only in a sub-figure letter or a parameter value; they enter the averages as one plot, the average of their panels (Section~\ref{sec:averaging}). The subsample of Section~\ref{sec:separated} allows us to ask which units of the analysis behave as independent: single plots, plot groups, or whole publications. Two questions are kept apart.

First, whether the two points of a plot are independent. Using only the first plot of every plot group, so that no averaging is involved, the binomial uncertainty of the $1\sigma$ fraction is \uncBinomAAtlas\% for ATLAS and \uncBinomACms\% for CMS. Resampling plots gives nearly the same, \uncPlotsAAtlas\% and \uncPlotsACms\%: the points of a plot behave as nearly independent.

Second, whether plot groups and publications are. With the groups averaged as in the analysis, independent points would give \uncBinomAtlas\% (ATLAS) and \uncBinomCms\% (CMS); resampling plot groups gives \uncGroupsAtlas\% and \uncGroupsCms\%, and resampling publications \uncPubsAtlas\% and \uncPubsCms\%. We test directly whether different plot groups of the same publication are alike. The weighted variance of the publications' $1\sigma$ fractions, each publication weighted by its size (Section~\ref{sec:averaging}), is compared with its distribution when the plot groups are reassigned to publications at random (\clusterPerm{} reassignments, keeping each publication's number of groups). The test is one-sided: the $p$-value is the fraction of reassignments with a weighted variance at least as large as the observed one (if none reaches it, $p \le 1/(N+1)$ for $N$ reassignments). For both experiments the observed weighted variance is larger than in nearly every reassignment ($p \clusterPAtlas$ for ATLAS with \clusterGroupsAtlas{} groups in \clusterPubsAtlas{} publications, $p \clusterPCms$ for CMS with \clusterGroupsCms{} groups in \clusterPubsCms{} publications). Groups of the same publication may resemble each other because they share data, background estimates and systematic uncertainties, or because publications differ in their coverage; the test cannot distinguish the two. Either way the publication, not the plot or the plot group, is the independent unit, and we quote bootstrap uncertainties over publications throughout. We generated 1000 toy data sets with the real plots, groups and publications, but independent points. In them, the binomial and bootstrap uncertainties agree with the spread of the fraction over the toys within 8\%. The real bootstrap uncertainties exceed those of nearly all toys at the levels of publications and plot groups; only for CMS with the separated points do a few reach them, 1\% of the toys at the level of publications and 4\% at the level of plot groups. Neither the structure of the data nor the code creates these differences.

The bootstrap treats publications as independent. Publications are not isolated, however: many cite earlier analyses of the same experiment, whose methods, background estimates or statistical treatment they may share. Using the citation links between the publications of our collection, retrieved from INSPIRE, we checked in two ways whether such links make publications alike in their coverage. Among the analyzed publications there are \citePairsAtlas{} citation links between \citeLinkedAtlas{} ATLAS publications and \citePairsCms{} links between \citeLinkedCms{} CMS publications.

The mean squared difference between the $1\sigma$ fractions of linked publications was compared with its distribution when the fractions are reassigned to the publications at random, keeping the citation links (20000 reassignments; one-sided, since similar linked publications would give a small difference). Linked publications are not significantly more alike than random pairs: $p \citePAllAtlas$ for ATLAS and $p \citePAllCms$ for CMS ($p \citePSepAtlas$ and $p \citePSepCms$ for the maximally separated points).

In addition, we repeated the analysis on the largest subset of publications in which no publication cites another, chosen to keep as many plots as possible (a maximum-weight independent set of the citation graph). The results, given in Table~\ref{tab:citefree}, agree with those of the full collection; their larger uncertainties reflect the smaller number of publications.

\begin{table*}[tb]
\centering
\topcaption{Results for all publications and for the largest subset of publications in which no publication cites another (column scan, bootstrap uncertainties).}\label{tab:citefree}
% generated by analysis/papertables.py from base git 2c7a2e8, 2026-11-03; do not edit
\begin{tabular}{lcccc}
\toprule
 & \multicolumn{2}{c}{CMS} & \multicolumn{2}{c}{ATLAS} \\
\cmidrule(lr){2-3}\cmidrule(lr){4-5}
 & All & Citation-free & All & Citation-free \\
\midrule
Publications & 27 & 19 & 28 & 16 \\
Plots & 205 & 165 & 168 & 115 \\
Raw $1\sigma$ (\%) & $69.5 \pm 3.2$ & $69.5 \pm 3.9$ & $81.8 \pm 5.9$ & $82.3 \pm 7.5$ \\
Raw $2\sigma$ (\%) & $95.8 \pm 1.2$ & $96.0 \pm 1.3$ & $99.5 \pm 0.3$ & $99.6 \pm 0.4$ \\
Norm. $1\sigma$ (\%) & $66.1 \pm 4.4$ & $66.2 \pm 5.0$ & $73.3 \pm 7.6$ & $69.3 \pm 10.1$ \\
Norm. $2\sigma$ (\%) & $95.3 \pm 1.9$ & $96.4 \pm 1.4$ & $99.2 \pm 0.4$ & $99.1 \pm 0.5$ \\
\bottomrule
\end{tabular}

\end{table*}

\subsection{Systematic Variations}\label{sec:cutoffs}

We quantify the systematic uncertainty associated with the choices of the analysis, including the selection cutoffs of Section~\ref{sec:thresholds}, by repeating the analysis with each choice varied. Table~\ref{tab:systematics} gives the results for the present collection. For each variation, it lists the fraction of points within the $1\sigma$ band, with its distance from the nominal 68.3\% in bootstrap standard deviations, and the fraction within the $2\sigma$ band. For the latter no such distance is given: these fractions lie close to 100\%, where the bootstrap distributions are strongly asymmetric.

The variations concern seven choices: (i)~the tolerance for matching band boundaries to an observed point: fixed at $\pm 1$~SVG unit for every plot as in the 2019 study instead of being scaled with the plotting frame, exact matching, and $\pm 0.5$ and $\pm 2$ units; (ii)~the column check of Section~\ref{sec:classification}; (iii)~the treatment of columns with nearly coinciding boundaries, discarded as in the 2019 study; (iv)~the merging of bins drawn as several steps; (v)~the crowding criterion, measured on the whole observed curve, and with thresholds of 90\% and 99\% instead of 95\%; (vi)~the exclusion of the looping and binned plots; and (vii)~all criteria of the 2019 study together. For the polygon method, points on a band boundary are counted entirely in the rarer or in the commoner of the two classes instead of half in each, and the distance within which a point counts as lying on a boundary is raised from 0.01 to 0.1 units.

For CMS, no variation of the column scan moves a fraction by more than 1.4 percentage points, and all remain consistent with the nominal values. For ATLAS, the normalized fractions move by at most 1.8 points. The raw $1\sigma$ fraction ranges from 77.7\% to 84.2\%, and stays between 1.6 and 3.5 standard deviations above the nominal value. Two variations stand out. Exact matching reduces the number of usable points, mostly for CMS, and moves the ATLAS raw $1\sigma$ fraction to 3.5 standard deviations above nominal. A crowding threshold of 99\% keeps most of the plots of arXiv:2503.16213, whose densely drawn curves contribute several thousand points. In raw averaging they then dominate the ATLAS result, whose $2\sigma$ fraction drops to 92.0\%; normalized averaging, which counts each plot group once, is hardly affected. This illustrates why the crowding criterion is needed. The crowding criterion measured on the whole observed curve, and the 90\% threshold, both exclude in addition one plot of arXiv:2201.07045, which lowers the ATLAS raw $1\sigma$ fraction by 3.6 points.

For the polygon method, the treatment of points on a band boundary changes the results by less than one percentage point. The distances between the observed points and the boundaries deciding their class are distributed smoothly, without an accumulation at the boundaries; a point therefore counts as lying on a boundary only within 0.01 units, which leaves genuine coincidences. A wider zone would set the number of such points arbitrarily: with 0.5 units, 17\% of the ATLAS points would fall into it, mostly from the closed contours of arXiv:2204.11988, whose observed contours run along band boundaries over long stretches. The column scan classifies points by their exact coordinates; for it, points within 0.01 units of a band boundary matched in their column are 0.1--0.6\% of the points, depending on the collection. The ends of the bands need care. Where the outline closes with a short vertical segment, or an edge has no vertex at the last value, the column holds more or fewer than two vertices of a color, and the column scan does not use the point. This concerns 64 first or last points of the present collection, about 0.8\% of its points. They lie, however, at the ends of the analyzed ranges, often at the highest masses, where the counts are lowest. The polygon method, which works with the band regions, includes them. In normalized averaging, a plot group weighs the same however few points it has. Leaving out the groups with fewer than 5 or 10 classified points changes the normalized fractions by at most 1.5 percentage points within the $1\sigma$ band and 0.7 within the $2\sigma$ band. Keeping the nine near copies of the present collection (Section~\ref{sec:averaging}) would change the results by at most 0.3 percentage points.

The 2019 comparison of Table~\ref{tab:revisit2019} changes by at most one percentage point in these variations (exact matching), mostly by a few tenths. The fixed $\pm 1$~SVG unit tolerance also admits the three plots of the CMS publication arXiv:1706.04260 (Figs.~3a, 3b and~4). Their frames are 10\% narrower than the standard one, and their band boundaries lie exactly one SVG unit from the observed points, so that the tolerance scaled with the frame misses them. The publication therefore enters the 2019 comparison only with the fixed tolerance.

\begin{table*}[tb]
\centering
\topcaption{Fractions of points within the bands (\%) for the present collection with each choice of the analysis varied (column scan unless stated; bootstrap uncertainties). In parentheses: distance of the $1\sigma$ fraction from the nominal 68.3\% in bootstrap standard deviations.}\label{tab:systematics}
% generated by analysis/systable.py
\small
\setlength{\tabcolsep}{4pt}
\begin{tabular}{lcccc}
\toprule
 & \multicolumn{2}{c}{Raw} & \multicolumn{2}{c}{Normalized} \\
\cmidrule(lr){2-3}\cmidrule(lr){4-5}
Variation & $1\sigma$ & $2\sigma$ & $1\sigma$ & $2\sigma$ \\
\midrule
\multicolumn{5}{l}{\textbf{ATLAS}} \\
Baseline & $81.8 \pm 5.9$ ($+2.3$) & $99.5$ & $73.3 \pm 7.6$ ($+0.7$) & $99.2$ \\
Fixed $\pm1$ unit tolerance & $81.8 \pm 5.9$ ($+2.3$) & $99.5$ & $73.3 \pm 7.6$ ($+0.7$) & $99.2$ \\
Column tolerance 0 (exact) & $84.2 \pm 4.6$ ($+3.5$) & $99.4$ & $71.5 \pm 8.8$ ($+0.4$) & $99.1$ \\
Column tolerance 0.5 & $81.9 \pm 5.8$ ($+2.3$) & $99.5$ & $73.2 \pm 7.6$ ($+0.7$) & $99.2$ \\
Column tolerance 2 & $81.8 \pm 5.9$ ($+2.3$) & $99.5$ & $73.3 \pm 7.6$ ($+0.7$) & $99.2$ \\
Column check off & $81.5 \pm 5.8$ ($+2.3$) & $99.3$ & $72.9 \pm 7.6$ ($+0.6$) & $98.9$ \\
Separation threshold as in 2019 & $81.5 \pm 6.0$ ($+2.2$) & $99.6$ & $73.1 \pm 7.7$ ($+0.6$) & $99.5$ \\
Step join off & $81.6 \pm 6.1$ ($+2.2$) & $99.5$ & $73.0 \pm 7.8$ ($+0.6$) & $99.2$ \\
Crowding on the whole curve & $78.2 \pm 6.1$ ($+1.6$) & $99.4$ & $73.0 \pm 7.7$ ($+0.6$) & $99.2$ \\
Crowding cut 90\% & $78.2 \pm 6.1$ ($+1.6$) & $99.4$ & $73.0 \pm 7.7$ ($+0.6$) & $99.2$ \\
Crowding cut 99\% & $77.7 \pm 4.0$ ($+2.3$) & $92.0$ & $73.4 \pm 7.3$ ($+0.7$) & $98.7$ \\
Without specialized samples & $81.3 \pm 6.2$ ($+2.1$) & $99.6$ & $72.2 \pm 8.4$ ($+0.5$) & $99.4$ \\
All 2019 rules & $81.5 \pm 6.0$ ($+2.2$) & $99.6$ & $73.1 \pm 7.7$ ($+0.6$) & $99.5$ \\
\cmidrule(lr){1-5}
Baseline, polygon method & $84.6 \pm 4.7$ ($+3.4$) & $97.6$ & $71.2 \pm 7.8$ ($+0.4$) & $98.8$ \\
Edge tolerance 0.1 & $84.5 \pm 4.8$ ($+3.4$) & $98.1$ & $71.2 \pm 7.8$ ($+0.4$) & $98.9$ \\
Edge points to the rarer class & $84.0 \pm 4.8$ ($+3.3$) & $97.3$ & $71.0 \pm 7.8$ ($+0.3$) & $98.6$ \\
Edge points to the commoner class & $85.2 \pm 4.7$ ($+3.6$) & $97.9$ & $71.5 \pm 7.8$ ($+0.4$) & $99.0$ \\
\midrule
\multicolumn{5}{l}{\textbf{CMS}} \\
Baseline & $69.5 \pm 3.2$ ($+0.4$) & $95.8$ & $66.1 \pm 4.4$ ($-0.5$) & $95.3$ \\
Fixed $\pm1$ unit tolerance & $69.5 \pm 3.2$ ($+0.4$) & $95.8$ & $66.1 \pm 4.4$ ($-0.5$) & $95.3$ \\
Column tolerance 0 (exact) & $70.9 \pm 3.5$ ($+0.7$) & $96.5$ & $66.5 \pm 4.5$ ($-0.4$) & $95.0$ \\
Column tolerance 0.5 & $69.9 \pm 3.2$ ($+0.5$) & $95.8$ & $67.1 \pm 4.4$ ($-0.3$) & $95.2$ \\
Column tolerance 2 & $69.5 \pm 3.2$ ($+0.4$) & $95.8$ & $66.1 \pm 4.4$ ($-0.5$) & $95.3$ \\
Column check off & $69.5 \pm 3.2$ ($+0.4$) & $95.8$ & $66.1 \pm 4.4$ ($-0.5$) & $95.3$ \\
Separation threshold as in 2019 & $69.0 \pm 3.2$ ($+0.2$) & $95.7$ & $66.2 \pm 4.4$ ($-0.5$) & $95.3$ \\
Step join off & $69.5 \pm 3.2$ ($+0.4$) & $95.8$ & $66.1 \pm 4.4$ ($-0.5$) & $95.3$ \\
Crowding on the whole curve & $69.5 \pm 3.2$ ($+0.4$) & $95.8$ & $66.1 \pm 4.4$ ($-0.5$) & $95.3$ \\
Crowding cut 90\% & $69.5 \pm 3.2$ ($+0.4$) & $95.8$ & $66.1 \pm 4.4$ ($-0.5$) & $95.3$ \\
Crowding cut 99\% & $69.5 \pm 3.2$ ($+0.4$) & $95.8$ & $66.1 \pm 4.4$ ($-0.5$) & $95.3$ \\
Without specialized samples & $68.8 \pm 3.5$ ($+0.1$) & $95.8$ & $65.2 \pm 5.0$ ($-0.6$) & $95.3$ \\
All 2019 rules & $69.0 \pm 3.2$ ($+0.2$) & $95.7$ & $66.2 \pm 4.4$ ($-0.5$) & $95.3$ \\
\cmidrule(lr){1-5}
Baseline, polygon method & $73.3 \pm 3.7$ ($+1.3$) & $96.2$ & $65.5 \pm 4.5$ ($-0.6$) & $95.2$ \\
Edge tolerance 0.1 & $73.3 \pm 3.7$ ($+1.4$) & $96.3$ & $65.6 \pm 4.5$ ($-0.6$) & $95.3$ \\
Edge points to the rarer class & $72.9 \pm 3.8$ ($+1.2$) & $96.0$ & $65.2 \pm 4.5$ ($-0.7$) & $94.9$ \\
Edge points to the commoner class & $73.7 \pm 3.7$ ($+1.5$) & $96.5$ & $65.9 \pm 4.5$ ($-0.5$) & $95.4$ \\
\bottomrule
\end{tabular}

\end{table*}

\section{Conclusions}\label{sec:conclusions}

We have extended our earlier meta-analysis of exclusion limits to the ATLAS and CMS searches for new physics published between 2020 and 2025, \nPapers{} publications with \nPlotsAtlas{} ATLAS and \nPlotsCms{} CMS exclusion plots, and revisited the publications of the 2019 study with the updated method. The limits are read directly from the vector graphics of the publications, and the fraction of observed points within the expected $1\sigma$ and $2\sigma$ bands is compared with the nominal 68.3\% and 95.4\%. The statistical uncertainties are estimated with a bootstrap over publications, which accounts for the correlations we find between the plots of a publication.

For CMS, the fractions agree with the nominal coverage: in normalized averaging, $\resNormOneCms\%$ of the points lie within the $1\sigma$ band and $\resNormTwoCms\%$ within the $2\sigma$ band. For ATLAS, the observed limits fall within the bands more often than nominal: $\resNormTwoAtlas\%$ within the $2\sigma$ band, while the $1\sigma$ fraction, $\resNormOneAtlas\%$, is compatible with the nominal value; in raw averaging the $1\sigma$ fraction, $\resRawOneAtlas\%$, lies \devRawOneAtlas{} standard deviations above it. The $2\sigma$ excess is also present where the band edges are well separated.

Observed limits that are biased, too strong or too weak, would lower the fraction within the bands. An excess points instead to bands that are wider than the spread of the observed limits, or to the low-count regime, where few events push the fraction up. Pseudo-experiments in simple models show that both can reach the size of the ATLAS excess, and that bands computed a posteriori, after a background-only fit to the data, add to it only where a signal region constrains a large uncertainty itself. That almost no ATLAS point lies beyond either $2\sigma$ edge favours bands that are too wide, from conservatively estimated uncertainties or a-posteriori bands, over the low-count regime alone. A related observation has been made within a single analysis. In the model-unspecific search of CMS, which compares data with the Standard Model expectation in several hundred event classes, slightly more classes than expected show only small deviations, and slightly fewer show moderate ones; the authors attribute this to a possible overestimation of systematic uncertainties~\cite[Sec.~8.3]{cms_music2021}. While their quantities are look-elsewhere-corrected $p$-values rather than exclusion limits, the direction of the effect is the one expected from conservatively estimated uncertainties. Which of these causes applies to a given analysis cannot be decided from its plots, nor whether the ATLAS excess reflects a common practice or particular analyses. Individual publications move the ATLAS fractions by up to about one standard deviation, but none of them determines the result.

The reading of the plots has been checked in several independent ways: against the numbers published in HEPData where available, against the markers with which many observed limits are drawn, against the colours of the rendered plots, and with a second classification method. The conclusions are stable under all variations of the choices made in the analysis, except a crowding threshold that readmits the densely drawn curves of arXiv:2503.16213 (Section~\ref{sec:cutoffs}). The HEPData records are valuable but not always reliable: in nine publications whose plots were all counted as available, the records omit two plots, tabulate three without both bands, and for five give numbers that cannot be right, such as $2\sigma$ offsets equal to the $1\sigma$ ones. Where the records are sound, they agree with the drawings to the precision of the vector graphics. A broader, consistent and checked availability of the numerical results would make studies like this one considerably simpler and allow them to be repeated routinely. Above all, the interpretation of the bands requires knowing how they were computed. A quick but methodical search of the texts of the publications in the present collection found none that states whether its expected limits and bands were computed a priori or a posteriori, although most state the type of limit and whether asymptotic formulae were used. We recommend that every publication with exclusion limits record, in its text or as HEPData keywords: (i)~the statistical tool and its version; (ii)~the type of limit; (iii)~whether asymptotic formulae or pseudo-experiments were used; and (iv)~whether the expected limits and their bands are computed a priori, a posteriori, or conditioned on the control regions only. Ideally the plot itself would show it, for instance with a legend entry naming the convention, since the same green and yellow bands currently stand for different quantities. This information matters for combinations and reinterpretations as much as for studies of coverage.

\backmatter

\bmhead{Acknowledgements}

V.E.\"O. acknowledges funding from the OpenMAPP project via the Scientific and Technological
Research Council of T\"urkiye (T\"UB\.{I}TAK, grant 223N165) under the EIC CHIST-ERA programme, and
gratefully acknowledges the software infrastructure provided by the Quantum4Edu project, funded by
T\"UB\.{I}TAK under grant 323K345.

We thank Emre \c{C}elebi of \.{I}stinye University and the ATLAS Statistics Forum for valuable
feedback on an earlier version of this analysis.

Large language models were used as assistants in this work. In developing the analysis code,
E.\c{S}. used GPT-4o (OpenAI) for assistance with specific programming questions. Claude Opus
5.5 (Anthropic) was used for reviewing and extending the analysis code, including the
implementation of the bootstrap uncertainty estimate and the automatic removal of duplicate
plots, for rerunning the analyses, and for editing the manuscript. It is worth mentioning that
had the same amount of work been performed by a student, that student would be considered
worthy of being a co-author on the paper. Regardless, all code and results have been reviewed
by the authors, who take full responsibility for the content of this paper.

\section*{Declarations}

\bmhead{Funding} V.E.\"O. acknowledges funding from the OpenMAPP project via T\"UB\.{I}TAK (grant 223N165) under the EIC CHIST-ERA programme.

\bmhead{Competing interests} V.E.\"Ozcan is a member of the ATLAS Collaboration and a co-author of the ATLAS publications analysed in this work. The authors have no financial interests to declare.

\bmhead{Data availability}
The exclusion plots analysed in this study (figures of ATLAS and CMS publications, redistributed under their Creative Commons licences), the lists of the analysed samples and the results are archived at Zenodo~\cite{zenodo_record}, \url{https://doi.org/10.5281/zenodo.24114098} (version v1; all versions: \url{https://doi.org/10.5281/zenodo.24114097}).

\bmhead{Code availability}
The analysis code is archived in the same record~\cite{zenodo_record}, \url{https://doi.org/10.5281/zenodo.24114098}, under the MIT licence.

\bmhead{Author contributions} E.\c{S}. collected the ATLAS and CMS publications and their exclusion plots, organised the samples and removed duplicate figures, and extended the analysis code of the 2019 study to the present collection, including the treatment of coordinate transformations, the selection of the observed limit, the matching tolerances and the methods for binned plots. She also prepared the tools for the citation links between publications and checked the availability of the analysed plots in HEPData. V.E.\"O. conceived the study, supervised the work, and directed the further development of the analysis, including the classification checks, the treatment of crowded and narrow-band plots, the statistical uncertainties, the independence and systematic studies, and the analysis pipeline (with the assistance of a large language model, as described in the Acknowledgements). Both authors interpreted the results, wrote and revised the manuscript, and approved the final version.

\bibliography{refs}

\clearpage
\end{document}